\documentclass[12pt]{article}

\usepackage[hang,flushmargin]{footmisc}
\usepackage[margin = 1in]{geometry}
\usepackage{algorithm}
\usepackage{algorithmicx}
\usepackage{algpseudocode}
\usepackage{blindtext}
\usepackage{graphicx}
\usepackage{natbib}
\usepackage{booktabs}
\usepackage{fancyhdr}
\usepackage{xr}
\usepackage{titling}
\usepackage{enumerate}
\usepackage{enumitem}

\usepackage{amsmath, amssymb, amsthm}
\usepackage{mathrsfs}
\usepackage{mathtools}
\usepackage{setspace}

\newtheorem{proposition}{Proposition}
\theoremstyle{remark}
\newtheorem{remark}{Remark}

\newcommand{\comment}[1]{}
\newcommand{\ZOIB}{\operatorname{ZOIB}}

\newcommand{\Dirichlet}{\operatorname{Dirichlet}}
\newcommand{\Uniform}{\operatorname{Uniform}}
\newcommand{\Normal}{\operatorname{Normal}}
\newcommand{\sX}{\mathscr X}
\newcommand{\sM}{\mathscr M}
\newcommand{\E}{\mathbb E}
\newcommand{\Beta}{\operatorname{Beta}}
\newcommand{\logit}{\operatorname{logit}}

\newcommand{\expit}{\operatorname{expit}}
\usepackage[dvipsnames]{xcolor}
\newcommand{\perpp}{\perp}

\usepackage[colorlinks, citecolor = BlueViolet]{hyperref}

\begin{document}

\title{Causal Mediation and Sensitivity Analysis for Mixed-Scale Data}

\author{
  Lexi Rene\thanks{Department of Statistics, Florida State University, email: \texttt{lor16b@my.fsu.edu}}, Antonio Linero\thanks{\parbox[t]{0.9\textwidth}{Department of Statistics and Data Science, University of Texas at Austin, \newline email: \texttt{antonio.linero@austin.utexas.edu}}}, and Elizabeth Slate\thanks{Department of Statistics, Florida State University, email: \texttt{eslate@fsu.edu}}
}

\maketitle

\begin{abstract}
  The goal of causal mediation analysis, often described within the potential outcomes framework, is to decompose the effect of an exposure on an outcome of interest along different causal pathways. Using the assumption of sequential ignorability to attain non-parametric identification, \citet{imai2010general} proposed a flexible approach to measuring mediation effects, focusing on parametric and semiparametric normal/Bernoulli models for the outcome and mediator. Less attention has been paid to the case where the outcome and/or mediator model are mixed-scale, ordinal, or otherwise fall outside the normal/Bernoulli setting. We develop a simple, but flexible, parametric modeling framework to accommodate the common situation where the responses are mixed continuous and binary, and apply it to a zero-one inflated beta model for the outcome and mediator. Applying our proposed methods to a publicly-available JOBS II dataset, we (i) argue for the need for non-normal models, (ii) show how to estimate both average and quantile mediation effects for boundary-censored data,  and (iii) show how to conduct a meaningful sensitivity analysis by introducing unidentified, scientifically meaningful, sensitivity parameters. 
\end{abstract}

\doublespacing

\section{Introduction}

\begin{figure}[t]
  \centering
  \includegraphics[width=.8\textwidth]{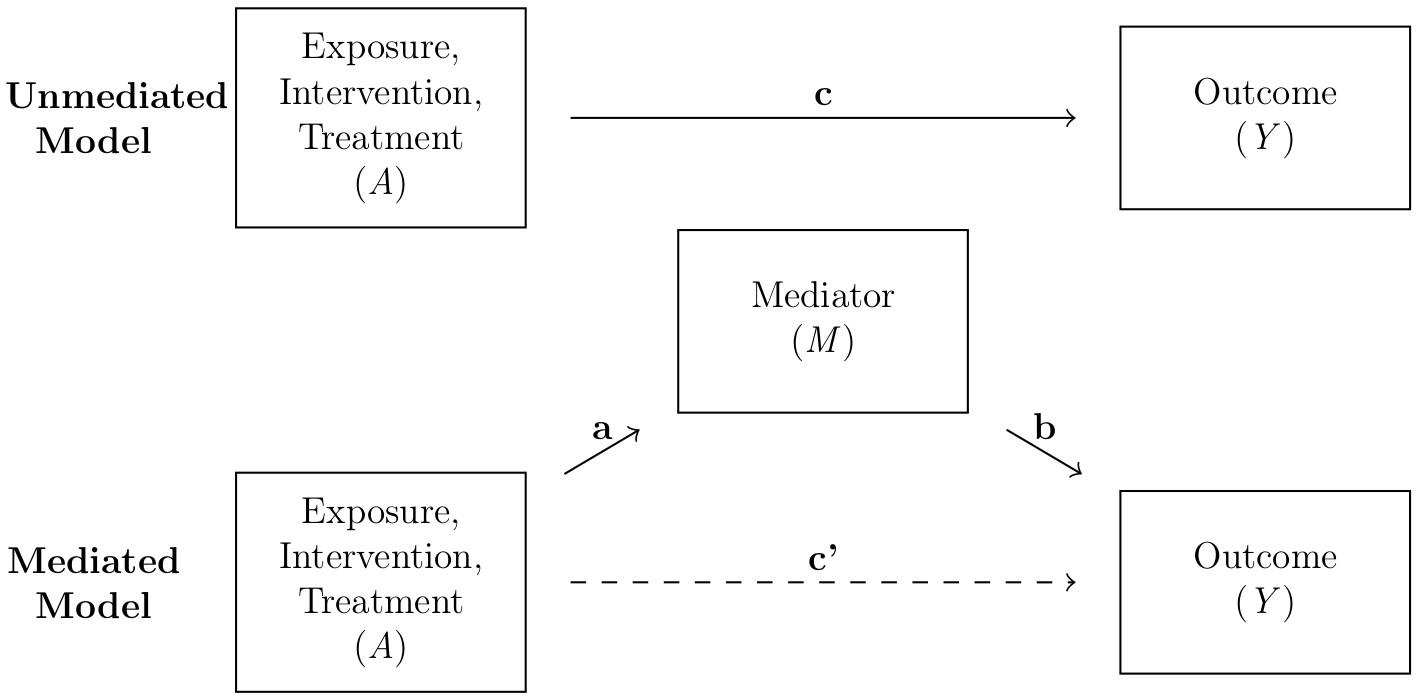}
  \caption{Schematic depiction of a causal structure with a single variable $M$ mediating the effect of the treatment $A$ on the outcome $Y$. Top: the causal structure ignoring the existence of the mediator. Bottom: the causal structure with the mediator included.}
  \label{fig:mediation}
\end{figure}

Mediation analysis is conducted across many scientific fields to understand the underlying mechanisms behind cause and effect relationships; examples include epidemiology, economics, and social science. Causal mediation analysis, often couched in the \emph{potential outcomes} framework \citep{rubin2004direct, imai2010general}, decomposes the effect of an exposure on the outcome along different causal pathways. 
A schematic depiction of a standard single-mediator model is given in Figure~\ref{fig:mediation}. In this diagram, $A$ denotes the exposure for the observational unit, $Y$ denotes the outcome, and $M$ denotes a mediator which may be on the causal pathway from the exposure to the outcome. When the mediator is accounted for in the relationship between $A$ and $Y$, we measure a \emph{direct effect} $(\mathbf c')$, while when the mediator is ignored we measure the \emph{total effect} $(\mathbf c)$. The \emph{indirect effect} of the exposure through its effect on the mediator uses pathways $\mathbf a$ and $\mathbf b$ to affect the outcome.

While most works on mediation analysis have focused on the case where the mediator and outcome are continuous/normal or Bernoulli distributed, in our experience it is common that one (or both) of the mediator or outcome will have a mixed-scale support. In this paper, we focus on the case where the mediator and outcome are \emph{mixed} continuous and discrete random variables; in particular, we assume that they have a continuous distribution on $(0,1)$ with mass at the boundary points $0$ and $1$. We argue that, particularly when taking a parametric Bayesian approach to estimation, it is important to adequately model the data, both for the purpose of reducing bias and to adequately account for uncertainty in effect estimation. To meet this challenge, we develop a general framework for performing causal mediation analysis with mixed-scale data. In principle, this framework can be used regardless of the model for the observed data, and we use a zero-one inflated beta regression model to illustrate.

\begin{figure}[t]
  \centering
  \includegraphics[width=1\textwidth]{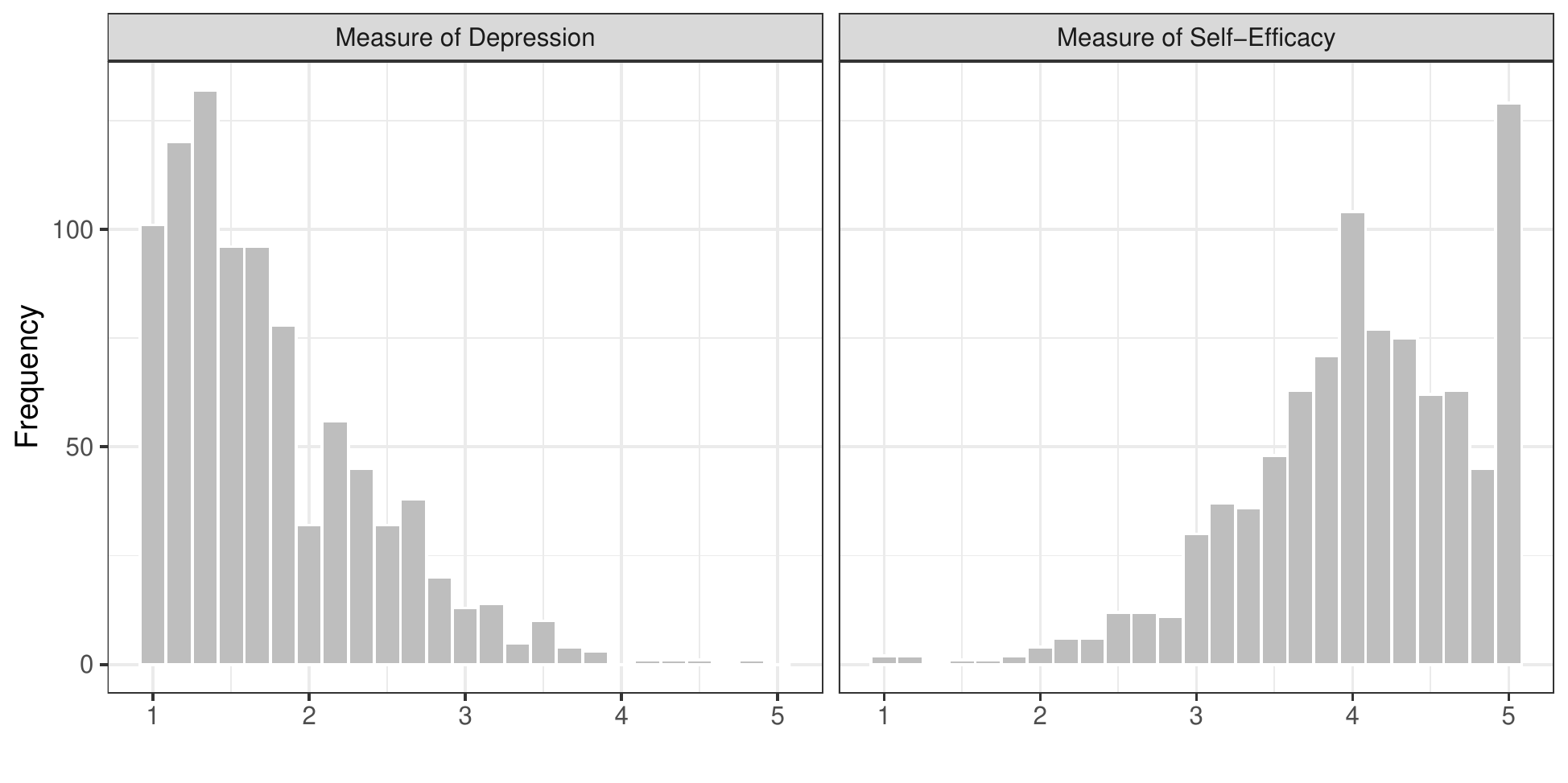}
  \caption{Empirical distribution of (left) measured depression level (right) measured job-search self-efficacy at the end of study and .}
  \label{fig:Figure-02-Lexi}
\end{figure}

For the sake of reproducibility we focus on the JOBS II study of \citet{vinokur1995impact}, for which a subset of data is available in the \texttt{mediation} package in \texttt{R}. A description of this dataset is given in Section~\ref{sec:jobs}. \citet{imai2010general} present several analyses of this dataset, essentially operating under the assumption that the mediator (a measure of self-confidence in finding a job) and the outcome (a measure of depression) are normally distributed. As shown in Figure~\ref{fig:Figure-02-Lexi}, however, it is apparent that neither the outcome nor the mediator is well-described by a normal distribution; both exhibit skewness and there is substantial mass at the boundary values of $1$ for depression and $5$ for self-confidence.

An additional challenge with mixed-scale models is assessing the sensitivity of inferences to untestable assumptions. As with most estimands in causal inference, it is well-known that the causal mediation effects are not identified on the basis of the observed data distribution alone, and can only be consistently estimated under additional (unfalsifiable) assumptions. A useful benchmark assumption is \emph{sequential ignorability} (SI, \citealp{imai2010general}), which essentially rules out the existence of unmeasured confounders. 
We found performing sensitivity analysis in the mixed-scale setting to be challenging, as to the best of our knowledge none of the existing proposals for sensitivity analysis can be applied directly. For example, the approaches proposed by \citet{imai2010general} are justified by a linear structural equation model (LSEM, \citealp{baron1986moderator}), which is not applicable in this setting. Similarly, the limited work with non-continuous or categorical data \citep{albert2015sensitivity, wang2012estimation} also does not apply directly to the mixed-scale setting. We develop a pair of widely-applicable sensitivity analysis strategies that accomplish the two goals of (i) assessing the extent to which our conclusions are driven by unmeasured confounding and (ii) neither imposing any additional restrictions on, nor adding information about, the distribution of the observed data. Our second goal is part of a recent trend in causal inference and missing data research of proposing sensitivity analyses that clearly and unambiguously separate the (parametric) assumptions used to model the observed data from the assumptions used to identify the causal effects of interest \citep{linero2017bayesian, LinDan2018, franks2019flexible, scharfstein2021semiparametric}.

\subsection{Review of Existing Methods}

The traditional approach to mediation analysis uses structural equation modeling (SEM) to quantify mediation; linear structural equation models (LSEMs) are particularly popular \citep{sem13}. 
However, LSEMs does not generalize easily to non-linear systems \citep{mackinnon93}. Additionally, \citet{imai2010general} makes the point that the identification assumptions used in LSEMs are inexorably tied to the choice of parametric model, stating: ``[because] the key identification assumption is stated in the context of a particular model, [it is] difficult to separate the limitations of research design from those of the specified statistical model.'' Motivated by this argument, \citet{imai2010general} proposed a more general approach to mediation analysis using a potential outcomes framework, introduced the nonparametric assumption of sequential ignorability to identify the effects, and showed that the single mediator LSEM is a special case of the potential outcomes framework which is valid as long as the linearity assumption holds.

There is a rich literature addressing the causal mediation problem from the semiparametric perspective. An emphasis in this literature is the development of methods that are both statistically efficient and \emph{multiply robust} in the sense that they produce consistent estimates even if one of several models required for estimation are misspecified \citep{tchetgen2012semiparametric, zheng2012targeted}. An advantage of these approaches is that one can easily use them with modern machine learning methods via cross-fitting \citep{farbmacher2020causal}. To the best of our knowledge, however, these methods have not been developed in the context of mixed-scale data. Bayesian nonparametric and semiparametric models based on infinite mixture models have also been proposed \citep{kim2016framework}, although not for mixed-scale data.

A variety of models have also extended beyond continuous/binary models for the mediator and outcome. These include models for zero-inflated count, survival, and ordinal data, as well as quantile regression models \citep{wang2012estimation, cheng2018mediation, vanderweele2011causal, imai2010general}.

\subsection{Contributions}

We make the following contributions in this paper. First, we describe how to implement the $g$-formula for computing both mean and quantile causal effects for generic mixed-scale models under the sequential ignorability assumption. Second, we illustrate these concepts using a zero-one inflated beta regression model and argue for its appropriateness on the benchmark JOBS II dataset. Third, we show how to conduct a principled sensitivity analysis to check the sensitivity of our conclusions to the untestable sequential ignorability assumption; the sensitivity parameters we introduce are designed to be unidentified, so that varying them does not affect the distribution of the observed data. We show how to introduce sensitivity parameters that are shifts of the mean on either a linear or logit scale; both of these approaches are very easy to incorporate into our models by post-processing the model fit. These mean-shift assumptions are weaker than the usual sequential ignorability assumption in that they only identify the mean of the potential outcomes rather than their whole distribution, but include the results of sequential ignorability as a special case.

We also present a flexible zero-one inflated beta model \citep{ospina2012general} for mediation analysis with boundary-censored data and show how to perform inference with this model. A Bayesian implementation of this model is given at \url{www.github.com/theodds/ZOIBMediation}. Our model is implemented in \texttt{Stan}, with the both average and quantile causal mediation effects computed using a Monte Carlo implementation of the $g$-formula. The Bayesian framework provides a straightforward approach to incorporating uncertainty in the sensitivity parameters through the use of informative priors, which can be elicited from subject-matter experts \citep{hogan2014}. In principle, however, one could also apply Frequentist inference using the nonparametric bootstrap \citep{linero2021simulation}.

\subsection{Outline}

In Section~\ref{sec:notation} we review the potential outcomes framework for mediation analysis and argue for the use of mixed-scale models on the JOBS II dataset. In Section~\ref{sec:zoib} we present our framework for causal mediation analysis, show how to compute the mean and quantile mediation effects, and develop our zero-one inflated beta regression model. In Section~\ref{sec:sensitivity-analysis} we present two alternative assumptions to sequential ignorability which allow for a sensitivity analysis, and show that these assumptions identify the average causal mediation effects. In Section~\ref{sec:illustrations} we illustrate our methodology on synthetic data and real data. We conclude in Section~\ref{sec:discussion} with a discussion. Proofs are in the appendix and some additional algorithms are given in the Supplementary Material. 

\section{Notation and Definitions of Causal Effects}
\label{sec:notation}

\subsection{JOBS II Dataset}
\label{sec:jobs}

For the sake of reproducibility, we motivate concepts and illustrate our methods on a subset of the JOBS II dataset \citep{vinokur1995impact} which is available publicly in the \texttt{mediation} package in \texttt{R}. The JOBS II data was used to evaluate the potential benefits of participation in a job-search skills seminar in southeastern Michigan. Subjects were recently unemployed adults during 1991-1993. Participants were pre-screened and classified according to their risk of depression and anxiety. High-risk participants, along with a random sample of low-risk participants, were invited to participate in the study. Prior to the seminar, questionnaires were sent out to the respondents. The questionnaires covered a range of topics about the respondent, including their loss of employment, the quality of work-life at their previous job, their health problems, and history of substance abuse. The primary baseline covariates, which we denote as $X_i$, include measures of education, income, race, marital status, age, sex, previous occupation, and level of economic hardship. The participants were randomly assigned to treatment and control groups. The treatment group, $A_i = 1$ were assigned to attend a seminar that taught participants job search skills and coping strategies for dealing with setbacks in the job hunt. The control group, $A_i = 0$, received a booklet of job search tips. Prior to measuring the outcome, but post-intervention, researchers measured an underlying mechanism in the relationship between the intervention and outcome. This \emph{mediator} was a continuous measure of job search self-efficacy, $M_i$. In this study two outcomes were measured: a continuous measure of depression using the Hopkins Symptom Checklist \citep{derogatis1974hopkins} and a binary variable for employment at the follow-up time. We will focus on the continuous measure of depression, $Y_i$.

Even for a benchmark dataset like the JOBS II data, which has been analyzed using LSEMs \citep{imai2010general}, there is overwhelming evidence that neither the outcome nor mediator are normally distributed. Observed values of the outcome and mediator, which are supported on $[1,5]$ and highly skewed, are displayed in Figure~\ref{fig:Figure-02-Lexi}, and it is apparent from the figure that the assumption of (say) a normally distributed error is untenable.


\subsection{The Potential Outcomes Framework}

Using the potential outcomes framework, the causal effect of the job training program can be defined as the difference between two potential outcomes. One potential outcome is realized if the subject participates in the training program and the other is realized if the subject does not.


Associated to the outcome and mediator are \emph{potential outcomes} that would have been observed had the treatment assignment been different. A potential outcome is defined as the outcome which would have been observed under an exposure a participant did not actually receive. We let $M_i(a)$ denote the value of the mediator had the treatment been assigned to the value $a$; in terms of the JOBS II study, this is the self-efficacy which would have been realized had the treatment for individual $i$ had been fixed at either receiving the treatment ($a=1$) or not ($a=0$). Similarly, we let $Y_i(a,m)$ denote the value of the outcome that would have been realized had the treatment for individual $i$ been fixed at $a$ and the mediator fixed at $m$; in terms of the JOBS II study, this is the depression level which would have been observed at a given level of self-efficacy under the two treatments.

We link the potential outcomes to the observed data through the \emph{consistency} assumption that $M_i = M_i(A_i)$ and $Y_i = Y_i\{A_i, M_i(A_i)\}$. The primary challenge in estimation of the mediation effects, i.e., the effects of changes of $M_i$ and $A_i$ on $Y_i$, lies in the fact that we cannot observe the counterfactual outcomes $Y_i\{A_i, M_i(1 - A_i)\}$, as this would require observing what would have happened under both $A_i = 1$ and $A_i = 0$.


\subsubsection{Sequential Ignorability}

As a starting point for identifying the causal effects of interest we will use the \emph{sequential ignorability} assumption of \citet{imai2010general}. For subject $i$, let $\sX$ be the support of the distribution of $X_i$ and let $\sM$ be the support of $M_i$. The sequential ignorability (SI) assumption imposes the following restrictions on the model parameterized by an unknown $\theta$:
\begin{align}
  \{Y_i(a', m), M_i(a)\} \perpp A_i &\mid X_i = x, \theta \tag{SI1} \qquad\qquad \text{and} \\
  Y_i(a', m) \perpp M_i(a) &\mid A_i = a, X_i = x, \theta \tag{SI2}
\end{align}
for all $a,a' = 0,1$, $x \in \sX$, and $m \in \sM$, where the expression $[U \perp V \mid W = w]$ means that $U$ is conditionally independent of $V$ given $W = w$. Additionally, we require the overlap condition (SI3) that $\Pr_\theta(A_i = a \mid X_i = x) > 0$ and $f_\theta\{M_i(a) = m \mid A_i = a, X_i = x\} > 0$. In words, SI1 states that, given the observed confounders, the treatment assignment is independent of the potential outcomes $Y_i(a',m)$ and $M_i(a)$; this will hold whenever the treatment assignment is randomized. On the other hand, SI2 states that the assignment of the mediator does not affect the outcome, given the observed treatment and pre-treatment covariates. Of the two assumptions, SI2 is generally the more problematic; for example, in the JOBS II study, the job-search self-efficacy is not randomized by study design, so that SI2 makes the uncheckable assertion that all common-causes of $M_i(a)$ and $Y_i(a,m)$ have been measured.

\subsubsection{Causal Mediation Effects}

We define the following causal mediation effects \citep{robins1992identifiability, pearl2001direct} using the JOBS II study for context. The \emph{natural indirect effect} (NIE), also called the \emph{causal mediation effect}, is defined for $a = 0,1$ as
\begin{align*}
  \delta_i(a) = Y_i\{a, M_i(1)\} - Y_i\{a, M_i(0)\}.
\end{align*}
In the JOBS II study, $\delta_i(1)$ is the difference between the two potential depression levels for subject $i$ if they participated in the job training seminar. The \emph{natural direct effect} (NDE) is defined for $a = 0,1$ as
\begin{align*}
  \zeta_i(a) = Y_i\{1, M_i(a)\} - Y_i\{0, M_i(a)\}.
\end{align*}
For example, in the JOBS II study, $\zeta_i(1)$ is the difference between the two potential depression levels for subject $i$ according to whether they participated in the job training seminar or not, under the assumption that their job search self-efficacy is held constant at the level which would have been observed if they had attended the seminar.

Because we cannot observe $Y_i\{a, M_i(a')\}$ when $a \ne a'$, we cannot directly observe either $\delta_i(a)$ or $\zeta_i(a)$. Nevertheless, under sequential ignorability we can estimate the \emph{average mediation effects}
\begin{align}
  \label{eq:mediation-effects}
  \delta(a) = \E_\theta[Y_i\{a, M_i(1)\} - Y_i\{a, M_i(0)\}]
  \quad\text{and}\quad
  \zeta(a) = \E_\theta[Y_i\{1, M_i(a)\} - Y_i\{0, M_i(a)\}].
\end{align}
The mediation effects $\delta(a)$ and $\zeta(a)$ decompose the \emph{average total effect} $\tau = \E_\theta[Y_i\{1, M_i(1)\} - Y_i\{0, M_i(0)\}]$ in the sense that $\delta(1) + \zeta(0) = \delta(0) + \zeta(1) = \tau$. The total effect is analogous to the usual average causal treatment effect (ATE) of the treatment assignment.

Under sequential ignorability, \citet{imai2010general} showed that the distribution of the potential outcomes $Y_i\{a, M_i(a')\}$ for any $a,a'$ is nonparametrically identified as
\begin{equation}
  \label{eq:si-identify}
  \begin{aligned}
  &f_\theta\left(
    Y_i\{a, M_i(a')\} = y \mid X_i = x
  \right)
  \\
  &\quad=
  \int_{\sM} f(Y_i = y \mid M_i = m, A_i = a, X_i = x) \,
            f(M_i = m \mid A_i = a', X_i = x) \
            dm
  \end{aligned}
\end{equation}
for all $x \in \sX$. The marginal distribution of $Y_i\{a, M_i(a')\}$ is then $\int f_\theta(Y_i\{a, M_i(a')\} = y \mid X_i = x) \ F_X(dx)$ so that it (along with the average direct and indirect effects) is also identified. While there usually will not be a simple analytical expression for \eqref{eq:si-identify}, it is nevertheless easy to approximate \eqref{eq:si-identify} using Monte Carlo integration; this approach was popularized by \citet{robins1986} as a tool to implement the $g$-formula in causal inference.

While the average causal mediation effects are the most commonly studied, one may also be interested in causal effects on other aspects of the distribution of the outcome. Let $Q_{q}(Z)$ denote the ${q}^{\text{th}}$ quantile of a random variable $Z$. Then the \emph{quantile mediation effects} at the quantile ${q}$ are
\begin{align}
    \label{eq:quantile}
    \begin{split}
    \delta_{q}(a) &= Q_{q}[Y_i\{a, M_i(1)\}] - Q_{q}[Y_i\{a, M_i(0)\}]
    \qquad\text{and}\qquad\\
    \zeta_{q}(a) &= Q_{q}[Y_i\{1, M_i(a)\}] - Q_{q}[Y_i\{0, M_i(a)\}].
    \end{split}
\end{align}
Because \eqref{eq:si-identify} fully identifies the distribution of $Y_i\{a,M_i(a')\}$, the quantile mediation effects are also identified under SI.

\section{Observed Data Models for Zero-One Inflated Data}
\label{sec:zoib}

Estimating the causal mediation effects under SI requires only that we estimate the distribution of the observed data. Without loss of generality, we assume that $Y_i$ and $M_i$ can be rescaled to lie in the interval $[0,1]$; for the JOBS II dataset, this can be done with the transformations $Y_i \gets (Y_i - 1) / 4$ and $M_i \gets (M_i - 1) / 4$, as the measures of depression and self-efficacy were measured on a scale from 1 to 5. 

\begin{figure}
    \centering
    \includegraphics[width=1\textwidth]{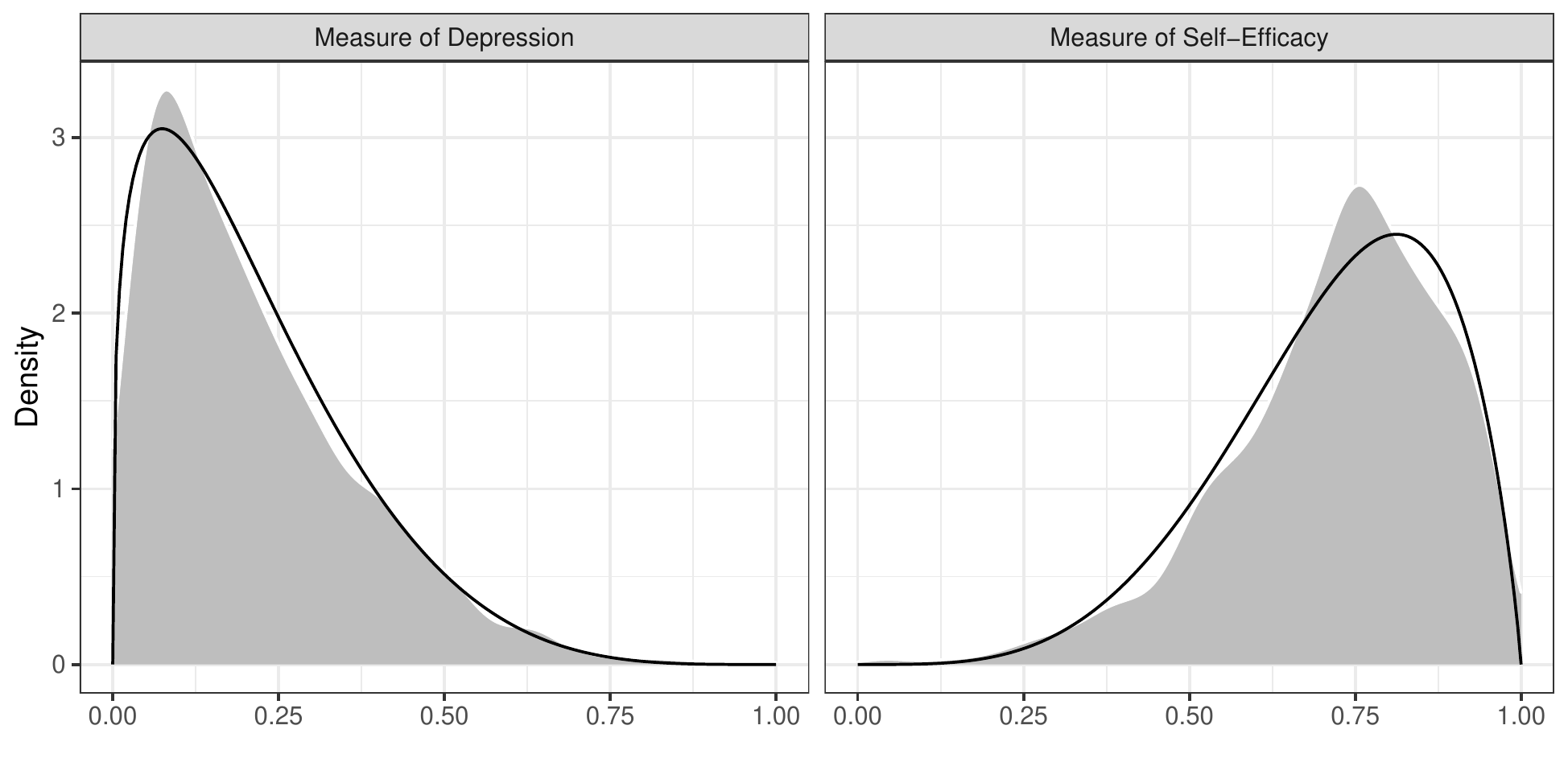}
    \caption{Kernel density estimate (gray) and fitted beta distribution (solid black) of the distribution of $Y_i$ and $M_i$ conditional on $Y_i \notin \{0,1\}$ and $M_i \notin \{0,1\}$.}
    \label{fig:Beta-Figs}
\end{figure}

A flexible distribution for zero-one inflated data on $[0,1]$ is the \emph{zero-one inflated beta} (ZOIB) distribution, which we denote by $\ZOIB(\alpha,\gamma,\mu,\phi)$. If $Z \sim \ZOIB(\alpha,\gamma,\mu,\phi)$ then $Z$ is a mixed discrete-continuous random variable such that
\begin{align}
  \label{eq:zoib}
  \Pr(Z = 0) = \alpha,
  \ \ 
  \Pr(Z = 1 \mid Z \ne 0) = \gamma,
  \ \ \text{and} \ \
  [Z \mid Z \notin \{0,1\}] \sim \Beta\{\mu \phi, (1 - \mu)\phi\}.
\end{align}
The parameterization of the beta distribution in \eqref{eq:zoib} is chosen so that $\mu$ is the mean of the beta distribution, i.e., $\E_\theta (Z \mid Z \notin\{0,1\}) = \mu$. The mean of the $\ZOIB(\alpha,\gamma,\mu,\phi)$ distribution is given by
\begin{align}
  \label{eq:zoib-mean}
  \E_\theta(Z)
  =
  (1-\alpha)\gamma + (1-\alpha)(1-\gamma)\mu.
\end{align}
Figure~\ref{fig:Beta-Figs} shows that the beta distribution is effective at modeling the shape of the data for the continuous part of the JOBS II data, while the parameters $\alpha$ and $\gamma$ allow for an increased chance of observing the boundary values $Z_i = 1$ and $Z_i = 0$.

Our ZOIB model assumes that $[Y_i \mid M_i = m, A_i = a, X_i = x] \sim \ZOIB(\alpha_i^Y, \gamma_i^Y, \mu_i^Y, \phi_i^Y)$ and $[M_i \mid A_i = a, X_i = x] \sim \ZOIB(\alpha_i^M, \gamma_i^M, \mu_i^M, \phi_i^M)$. We model the parameters of these ZOIB distributions with generalized linear models of the form
\begin{equation}
  \label{eq:glms}
  \begin{aligned}
  \logit(\alpha_i^Y) &= (X_i, M_i)^\top \beta_\alpha^Y(A_i), &
  \logit(\gamma_i^Y) &= (X_i, M_i)^\top \beta_\gamma^Y(A_i), \\
  \logit(\mu_i^Y)    &= (X_i, M_i)^\top \beta_\mu^Y(A_i), &
  \log(\phi_i^Y)     &= (X_i, M_i)^\top \beta_\phi^Y(A_i).
  \end{aligned}
\end{equation}
The dependence of the $\beta$'s on $A_i$ is included to allow for heterogeneous effects of the covariates and mediator; the homogeneous model is included as a special case where only the intercept varies with $A_i$. Similar models are specified for $(\alpha_i^M, \gamma_i^M, \mu_i^M, \phi_i^M)$. As a default, all of the regression coefficients are given flat $\Normal(0,\tau^2)$ priors, where $\tau^2$ is taken to be large after centering and scaling the covariates $X_i$ (except for the intercept) to have mean $0$ and standard deviation $1$.

To estimate the mediation effects we also require a model for the distribution $F_X$ of the covariates. As a default, we assume that $F_X$ is discretely supported on the observed values of the $X_i$'s, i.e., $\Pr_\theta(X_i = x_j) = \omega_j$ where $(x_1, \ldots, x_N)$ are the observed values of the covariates. We then specify an improper \emph{Bayesian bootstrap} \citep{rubin1981bayesian} prior for $\omega = (\omega_1, \ldots, \omega_N)$, i.e., $\pi(\omega) = \prod_{i} \omega_i^{-1}$. After observing the data, the posterior distribution of $\omega$ is $\Dirichlet(1,\ldots,1)$, and so can be sampled exactly. Specifying a Bayesian bootstrap prior for $F_X$ avoids the notoriously difficult task of estimating $F_X$ via density estimation, which has been shown in other settings to result in improved theoretical properties of Bayesian causal inference methods \citep{ray2020semiparametric}.

\section{Posterior Computation and Inference}
\label{sec:posterior-computation-and-inference}

We divide inference into two steps:
\begin{enumerate}
\item
  Draw a set of approximate samples $\theta_1, \ldots, \theta_B$ from the posterior distribution of $\theta$.
\item
  For each $\theta_b$, compute $\delta(a), \zeta(a)$, and $\tau$, yielding approximate samples from the posterior distribution for these mediation effects.
\end{enumerate}
For the first step we use the probabilistic programming language \texttt{Stan}, which implements an adaptive version of Hamiltonian Monte Carlo (HMC) to sample $\theta_b$'s \citep{carpenter2016stan}. The sole exception to this sampling scheme is that we sample $\omega \sim \Dirichlet(1,\ldots,1)$ directly from the posterior distribution.

\subsection{Average Mediation Effects}

Due to the nonlinearities of the ZOIB model, the mediation effects $\delta(a), \zeta(a),$ and $\tau$ are not available in closed form and must be approximated. To compute the mediation effects we use a Monte Carlo implementation of the $g$-formula. The idea is to note that, because \eqref{eq:si-identify} identifies the distribution of $Y_i\{a, M^\star_i(a')\}$ for all $a, a'$, we can simulate $K$ new realizations $Y_i^\star\{a, M^\star_i(a')\}$ for $i=1,\ldots,N$ from the model, in which case
\begin{align}
  \label{eq:estimates-naive}
  \begin{split}
  \delta(a) & \approx K^{-1}\sum_{i,k}\omega_i [Y_{ik}^{\star}\{a,M_{ik}^{\star}(1)\}-Y_{ik}^{\star}\{a,M_{ik}^{\star}(0)\}] \qquad\text{and}\\
  \zeta(a) & \approx K^{-1}\sum_{i,k}\omega_i[Y_{i}^{\star}\{1,M_{ik}^{\star}(a)\}-Y_{ik}^{\star}\{0,M_{ik}^{\star}(a)\}] 
  \end{split}
\end{align}
are unbiased estimators of $\delta(a)$ and $\zeta(a)$. The approximations in \eqref{eq:estimates-naive} are less efficient than using the true values $\delta(a)$ and $\zeta(a)$ because they contain Monte Carlo error, but are \emph{conservative} in the sense that using them results in valid inference. The approximations can be improved by using various tricks to eliminate the Monte Carlo error. One improvement is to notice that we can decrease the variance of \eqref{eq:estimates-naive} by replacing $Y^\star_{ik}\{a, M^\star_{ik}(a')\}$ with the conditional expectation $\E_\theta[Y^\star_{ik}\{a, M^\star_{ik}(a')\} \mid M_{ik}^\star(a'), X_i]$; for the ZOIB model, this is given by $(1 - \alpha^Y_{ik}) \gamma^Y_{ik} + (1-\alpha^Y_{ik})(1 - \gamma^Y_{ik}) \mu^Y_{ik}$ where $\alpha^Y_{ik}, \gamma_{ik}^Y$ and $\mu_{ik}^Y$ are given by \eqref{eq:glms} with $A_i$ evaluated at $a$ and $M_i$ evaluated at $M^\star_{ik}(a')$.

\begin{algorithm}[t]
  \caption{Monte Carlo $g$-formula Estimate of Average Causal Effects \label{alg:gc}}
  \textbf{Input:} $\theta, K, \{X_i\}_{i=1}^N$
  \begin{algorithmic}[1]
    \State{Sample $\omega \sim \Dirichlet(1,\ldots,1)$}
    \For{$k = 1, \ldots, K$ and $i = 1,\ldots,N$}
      \State{
        Sample $U_{ik} \sim \Uniform(0,1)$
      }
      \State{
        Set $M^\star_{ik}(a) = F_M^{-}(U_{ik} \mid A = a, X = X_i,\theta)$ for $a = 0,1$.
      }
      \State{
        Set
        $
        Y^\star_{ik}\{a, M^\star_{ik}(a')\} 
        = 
        \E_\theta\{Y_i \mid M_i = M_{ik}^\star(a'), A_i = a, X_i\}
        $ 
        for $a, a' = 0,1$.
      }
    \EndFor
    \State{
      Approximate $\delta(a),\zeta(a), \tau$ with $\widehat \delta(a), \widehat \zeta(a), \widehat \tau$, respectively, where
      \begin{align*}
        \widehat \delta(a) & = K^{-1}\sum_{i,k}\omega_i [Y_{i}^{\star}\{a,M_{i}^{\star}(1)\}-Y_{i}^{\star}\{a,M_{i}^{\star}(0)\}] \\
        \widehat \zeta(a) & = K^{-1}\sum_{i,k}\omega_i[Y_{i}^{\star}\{1,M_{i}^{\star}(a)\}-Y_{i}^{\star}\{0,M_{i}^{\star}(a)\}]  \\
        \widehat \tau & = K^{-1}\sum_{i,k}\omega_i[Y_{i}^{\star}\{1,M_{i}^{\star}(1)\}-Y_{i}^{\star}\{0,M_{i}^{\star}(0)\}]
      \end{align*}   
      \hspace*{.4em}for $a = 0,1$
    }
    \State{\Return{$\{\widehat \delta(0), \widehat \delta(1), \widehat \zeta(0), \widehat \zeta(1), \widehat \tau\}$}}
  \end{algorithmic}
\end{algorithm}

The Monte Carlo integration strategy is summarized in Algorithm~\ref{alg:gc}, and it applies to any model. In Algorithm~\ref{alg:zoib} we give the special case of our ZOIB regression models. In this algorithm, $F^-_M(u \mid A_i = a, X_i = x) = \inf \{m : F_M(m \mid A_i = a, X_i = x) \ge u\}$ denotes the \emph{generalized inverse} of the cumulative distribution function of $[M_i \mid A_i = a, X_i = x]$.

\begin{algorithm}
  \caption{Monte Carlo $g$-formula for ZOIB Model\label{alg:zoib}}
  \textbf{Input:} $\theta, K, \{X_i\}_{i=1}^N$
  \begin{algorithmic}[1]
    \State{Sample $\omega \sim \Dirichlet(1,\ldots,1)$}
    \For{$k = 1,\ldots,K$ and $i=1,\ldots,N$}
        \State Sample $U \sim \Uniform(0,1)$
        \For{$a = 0,1$}
          \State{Compute \ \parbox[t]{.7\textwidth}{
            $\logit \alpha^M \gets X_i^\top \beta^M_\alpha(a)$,
            $\logit \gamma^M \gets X_i^\top \beta^M_\gamma(a)$,\newline
            $\logit \mu^M \gets X_i^\top \beta^M_\mu(a)$,
            $\log \phi^M \gets X_i^\top \beta^M_\phi(a)$.
          }
          }
          \State{
            Set
            \begin{align*}
              M_{ik}^\star(a)
              \gets
              \begin{cases}
                0 \qquad & \text{if $U < \alpha^M$}, \\
                1 \qquad & \text{if $U > 1 - (1-\alpha^M)\gamma^M$}, \\
                F_{\Beta}^{-1}\{U' \mid \mu^M \phi^M, (1-\mu^M)\phi^M\} \qquad & \text{otherwise}
              \end{cases}
            \end{align*}
            \hspace*{4.8em}where $U' = (U - \alpha^M) / [(1 - \alpha^M)(1 - \gamma^M)]$.
          }
          \For{$a' = 0,1$}
            \State{Compute
                \parbox[t]{.72\textwidth}{
                $\logit \alpha^Y \gets (X_i, M_{ik}^\star(a))^\top \beta^Y_\alpha(a')$, 
                $\logit \gamma^Y \gets (X_i, M_{ik}^\star(a))^\top \beta^Y_\gamma(a')$,\newline
                $\logit \mu^Y \gets (X_i, M_{ik}^\star(a))^\top \beta^Y_\mu(a')$,
                $\log \phi^Y \gets (X_i, M_{ik}^\star(a))^\top  \beta^Y_\phi(a')$.
                }
            }
            \State{Set
            \begin{math}
               Y_{ik}^\star(a', M_{ik}^\star(a))
              \gets
              (1-\alpha^Y) \gamma^Y + (1-\alpha^Y)(1-\gamma^Y)\mu.
            \end{math}
            }
          \EndFor
        \EndFor
    \EndFor
    \comment{
    \State{Approximate $\delta(a),\zeta(a), \tau$ with
      \begin{align*}
        \delta(a) & \approx K^{-1}\sum_{i,k}\omega_i [Y_{ik}^{\star}\{a,M_{ik}^{\star}(1)\}-Y_{ik}^{\star}\{a,M_{ik}^{\star}(0)\}] \\
        \zeta(a) & \approx K^{-1}\sum_{i,k}\omega_i[Y_{ik}^{\star}\{1,M_{ik}^{\star}(a)\}-Y_{ik}^{\star}\{0,M_{ik}^{\star}(a)\}]  \\
        \tau & \approx K^{-1}\sum_{i,k}\omega_i[Y_{ik}^{\star}\{1,M_{ik}^{\star}(1)\}-Y_{ik}^{\star}\{0,M_{ik}^{\star}(0)\}]
      \end{align*}   
      \hspace*{0em}for $a = 0,1$
    }
    \State{\Return{$\{\delta(0), \delta(1), \zeta(0), \zeta(1), \tau\}$}}
    }
    \State{Approximate $\delta(a),\zeta(a), \tau$ with $\widehat \delta(a), \widehat \zeta(a), \widehat \tau$, respectively, where
      \begin{align*}
        \widehat \delta(a) & = K^{-1}\sum_{i,k}\omega_i [Y_{ik}^{\star}\{a,M_{ik}^{\star}(1)\}-Y_{ik}^{\star}\{a,M_{ik}^{\star}(0)\}] \\
        \widehat \zeta(a) & = K^{-1}\sum_{i,k}\omega_i[Y_{ik}^{\star}\{1,M_{ik}^{\star}(a)\}-Y_{ik}^{\star}\{0,M_{ik}^{\star}(a)\}]  \\
        \widehat \tau & = K^{-1}\sum_{i,k}\omega_i[Y_{ik}^{\star}\{1,M_{ik}^{\star}(1)\}-Y_{ik}^{\star}\{0,M_{ik}^{\star}(0)\}]
      \end{align*}   
      \hspace*{0em}for $a = 0,1$
    }
    \State{\Return{$\widehat \{\delta(0), \widehat \delta(1), \widehat \zeta(0), \widehat \zeta(1), \widehat \tau \}$}}
  \end{algorithmic}
\end{algorithm}

\subsection{Quantile Mediation Effects}

Equation \eqref{eq:si-identify} can also be used to form a Monte Carlo estimate of the quantile mediation effects, although the implementation is somewhat more subtle. If we have a sample of $Y_{ik}^\star\{a, M^\star_i(a')\}$'s from the marginal density $f_\theta(Y_i\{a, M_i(a')\} = y)$ then we can approximate its ${q}^{\text{th}}$ quantile as $Q_{q}(\mathbb F_{aa'})$ where $\mathbb F_{aa'}$ is the empirical distribution of our sample and $Q_{q}(F)$ is the ${q}^{\text{th}}$ quantile of $F$. We can then calculate \eqref{eq:quantile} using the approximation
\begin{align}
    \label{eq:quantile-approx}
    \delta_{q}(a) 
    \approx 
    Q_{q}(\mathbb F_{a1}) - Q_{q}(\mathbb F_{a0})
    \quad\text{and}\quad
    \zeta_{q}(a)
    \approx
    Q_{q}(\mathbb F_{1a}) - Q_{q}(\mathbb F_{0a}).
\end{align}
Note that for this to be valid we must sample the covariates $X_{ik}^\star$'s used to generate $M_{ik}^\star(a')$ and $Y^\star_{ik}\{a, M_{ik}^\star(a')\}$ according to $\omega_i$, rather than averaging over $\omega$ as in \eqref{eq:estimates-naive}. This results in higher Monte Carlo error in \eqref{eq:quantile-approx} than in \eqref{eq:estimates-naive}.

Reducing Monte Carlo error in \eqref{eq:quantile-approx} can also be done, although it requires different strategies; for example, it is no longer valid to replace $Y^\star_{ik}\{a, M_{ik}^\star(a')\}$ with its mean. One may take $K$ very large, but this can substantially increase computation time. Another trick is to construct the joint distribution of $\{Y_{ik}^\star\{a, M_{ik}^\star(a')\}: a, a' \in \{0,1\}\}$ in a way which makes the potential outcomes highly correlated. Interestingly, because \eqref{eq:quantile} depends only on the marginal distributions of the potential outcomes, it is invariant to our choice of joint distribution; hence, this does not actually imply any additional restrictions on the model. To ensure a strong dependence between the $Y_i^\star\{a, M^\star_i(a')\}$'s, we simulate $M_i^\star(a)$ and $Y^\star_i(a, m)$ to be \emph{comonotone} \citep{deelstra2011overview}, i.e., we simulate $U, V \sim \Uniform(0,1)$ and apply the probability integral transform to get $M_i^\star(a) = F^-_M(U \mid A_i = a, X_i)$ and $Y_i^\star\{a, M_i^\star(a')\} = F^-_Y(V \mid M_i = M_i^\star(a'), A_i = a, X_i)$ (note that the same $U$ and $V$ are used for different values of $a$ and $a'$). This, combined with taking $K$ to be modestly large (say, $K = 10$) is sufficient to effectively eliminate the Monte Carlo error.

\begin{algorithm}[t]
  \caption{Monte Carlo $g$-formula Estimate of Quantile Causal Effects\label{alg:gc-quantile}}
  \textbf{Input:} $\theta, K, \{X_i\}_{i=1}^N$
  \begin{algorithmic}[1]
    \State{Sample $\omega \sim \Dirichlet(1,\ldots,1)$}
    \For{$k = 1, \ldots, K$ and $i = 1,\ldots,N$}
      \State{
        Sample $X_{ik}^\star = X_j$ with probability $\omega_j$
      }
      \State{
        Sample $U_{ik}, V_{ik} \sim \Uniform(0,1)$
      }
      \State{
        Set $M^\star_{ik}(a) = F_M^{-}(U_{ik} \mid A = a, X = X^\star_{ik},\theta)$ for $a = 0,1$.
      }
      \State{
        Set $Y^\star_{ik}\{a, M^\star_{ik}(a')\} = F_Y^{-}(V_{ik} \mid M = M^\star_{ik}(a'), A = a, X = X^\star_{ik}, \theta)$ for $a, a' = 0,1$.
      }
    \EndFor
    \comment{
    \State{
      Approximate $\delta(a),\zeta(a), \tau$ with
      \begin{align*}
        \delta_{q}(a) &\approx Q_{q}a(\mathbb F_{a1}) - Q_{q}(\mathbb F_{a0}) \\
        \zeta_{q}(a) & \approx Q_{q}(\mathbb F_{1a}) - Q_{q}(\mathbb F_{0a}) \\
        \tau_{q} & \approx Q_{q}(\mathbb F_{11}) - Q_{q}(\mathbb F_{00})
      \end{align*}   
      \hspace*{.4em}for $a = 0,1$, where $\mathbb F_{aa'}$ is the empirical distribution of the $Y^\star_{ik}\{a, M^\star_{ik}(a')\}$'s.
    }
    \State{\Return{$\{\delta_{q}(0), \delta_{q}(1), \zeta_{q}(0), \zeta_{q}(1), \tau_{q} \}$}}
    }
    \State{
      Approximate $\delta_{q}(a),\zeta_{q}(a), \tau_{q}$ with $\widehat \delta_{q}(a), \widehat \zeta_{q}(a), \widehat \tau_{q}$, respectively, where
      \begin{align*}
        \widehat \delta_{q}(a) & = Q_{q}(\mathbb F_{a1}) - Q_{q}(\mathbb F_{a0}) \\
        \widehat \zeta_{q}(a) & = Q_{q}(\mathbb F_{1a}) - Q_{q}(\mathbb F_{0a}) \\
        \widehat \tau_{q} & = Q_{q}(\mathbb F_{11}) - Q_{q}(\mathbb F_{00})
      \end{align*}   
      \hspace*{.4em}for $a = 0,1$, where $\mathbb F_{aa'}$ is the empirical distribution of the $Y^\star_{ik}\{a, M^\star_{ik}(a')\}$'s.
    }
    \State{\Return{$\{\widehat \delta_{q}(0), \widehat \delta_{q}(1), \widehat \zeta_{q}(0), \widehat \zeta_{q}(1), \widehat \tau_{q} \}$}}
  \end{algorithmic}
\end{algorithm}

Our general algorithm for approximating the quantile mediation effects is given in Algorithm~\ref{alg:gc-quantile}, with the extension to the specific setting of the ZOIB model being derived in the same way Algorithm~\ref{alg:zoib} was derived from Algorithm~\ref{alg:gc}.


\subsection{Assessing the Monte Carlo Error}

\citet{linero2021simulation} introduced a method for computing and (in the case where the effects are approximately normal) correcting for the Monte Carlo error in the types of estimators we have proposed; code for implementing this is available at \url{www.github.com/theodds/AGC}. This approach requires $K > 1$ and, in all cases we have considered, the Monte Carlo error is negligible for $K = 2$. \citet{linero2021simulation} also shows that naive estimators that are not designed to eliminate Monte Carlo error can be very inefficient unless $K$ is taken rather large (say, $K \ge 10$).

\section{Sensitivity Analysis}
\label{sec:sensitivity-analysis}

Because SI is an untestable assumption, it is essential to assess the extent to which the conclusions of an SI-based analysis are sensitive to the existence of unmeasured confounders, i.e., SI2. Accordingly, we now present a framework for performing sensitivity analysis using the mixed-scale models we have developed to model the observed data. Without loss of generality, we assume that the data has been scaled so that both $Y_i(a,m)$ and $M_i(a)$ take values on $[0,1]$. As a guiding principle, we require that any sensitivity parameters be \emph{pure} sensitivity parameters in the sense that varying them does not alter the fit of the model to the data. This allows us to clearly separate the goodness-of-fit to the observed data from sensitivity to the SI assumption.

\subsection{Sensitivity on the Logit Scale}
\label{sec:sensitivity-on-the-logit-scale}

We propose an approach to sensitivity analysis that allows for dependence between $Y_i(a,m)$ and $M_i(a)$ even after accounting for $X_i$ and $A_i$. We replace assumption SI2 with the following two assumptions:
\begin{itemize}
    \item[SI2A] Conditional on $X_i$, the potential outcomes $M_i(0)$ and $M_i(1)$ are jointly distributed according to a Gaussian copula with correlation $\rho \in [0,1]$. More precisely, we have $M_i(a) = F_M^-\{\Phi(Z_{ia}) \mid A_i = a, X_i = x\}$ where $F_M^-(u \mid A_i = a, X_i = x)$ denotes the generalized inverse cdf of $M_i$ given $A_i = a$ and $X_i = x$, and $(Z_{i0}$, $Z_{i1})$ is jointly standard normal with correlation $\rho$.
    \item[SI2B] Conditional on $X_i$, $M_i(0)$, and $M_i(1)$, the mean of $Y_i(a,m)$ is given by 
    \begin{align*}
      \E_\theta\{Y_i(a,m) \mid M_i(a), M_i(a'), X_i\}
      =
      \expit[\logit r_y(m,a,x) + \lambda\{M_i(a) - m\}]
    \end{align*}
    where $r_y(m,a,x) = \E_\theta(Y_i \mid M_i = m, A_i = a, X_i = x)$.
\end{itemize}
SI2B has been chosen specifically so that it reproduces the inferences under SI2 when $\lambda = 0$ while leaving the sensitivity parameters $\lambda$ and $\rho$ unidentified so that they can be varied freely without changing the fit of the model to the data. The most closely related sensitivity analysis framework which we are aware of is the ``hybrid'' approach of \citet{albert2015sensitivity}, although this approach differs in two important ways: first, it is only applied for the identity link and so (for reasons outlined in Section~\ref{sec:sensitivity-on-the-linear-scale}) does not require specifying $\rho$; and second, it replaces the term $\lambda\{M_i(a) - m\}$ with a term of the form $\lambda(a - a')$, which is similarly designed to drop out of the distribution of the observed data.

The following proposition establishes that $\E_\theta[Y_i\{a,M_i(a')\}]$ is identified for all $a, a'$ so that the average causal mediation effects are also identified. A proof is given in Appendix~\ref{sec:proofs}. 

\begin{proposition}
  \label{prop:logit}
  Suppose that SI1, SI2A, SI2B, and SI3 hold. Then we have
  \begin{align*}
    &\E_\theta[Y_i\{a, M_i(a')\}]  
    \\&\qquad=
    \iiint
    \expit\{\widetilde r_y(m',a,x) + \lambda(m - m')\}
    \, 
    \Normal\{(z_0, z_1)^\top \mid (0,0)^\top, \Sigma\}
    \ dz_0 \ dz_1 \ F_X(dx)
  \end{align*}
  where $\widetilde r_y(m,a,x) = \logit r_y(m,a,x)$, $\Sigma = \begin{psmallmatrix}1 & \rho \\ \rho & 1\end{psmallmatrix}$, and we define $m' = F^-_M\{\Phi(z_{a'}) \mid A_i = a', X_i = x)$ and $m = F^-_M\{\Phi(z_a) \mid A_i = a, X_i = x\}$ in the integral.
\end{proposition}
As in Section~\ref{sec:posterior-computation-and-inference}, there is no analytical expression for $\E_\theta[Y_i\{a, M_i(a')\}]$, and hence we must resort to Monte Carlo integration. Fortunately, by noting that the approximation
\begin{align*}
    \E_\theta[Y_i\{a, M_i(a')\}]
    \approx
    K^{-1} \sum_{i,k} 
    \omega_i 
    \expit[\widetilde r_y\{M^\star_i(a'), a, X_i\} + \lambda\{M^\star_i(a) - M^\star_i(a')\}]
\end{align*}
is unbiased, it is straight-forward to modify Algorithm~\ref{alg:gc} to compute $\E_\theta[Y_i\{a, M_i(a')\}]$ under this assumption as well. A procedure for computing the mediation effects under SI2A and SI2B is given in Algorithm~\ref{alg:sensitivity} of the Supplementary Material. 

As there is no information in the data about the sensitivity parameters $\rho$ and $\lambda$, it is essential that they be interpretable so that plausible ranges for them can be elicited. To gain better intuition about the role of $\lambda$, suppose that we had instead posited the logistic regression model $\logit r_y(m,a,x) = \beta^Y_0 + x^\top \beta^Y_X + a \, \beta^Y_A + m \, \beta^Y_M$. In this case, we can rewrite
$$
  \logit \E_\theta\{Y_i(a,m) \mid M_i(a), M_i(a'), X_i\}
  =
  \beta^Y_0 + x^\top \beta^Y_X + a \, \beta^Y_A + m \, (\beta^Y_M - \lambda) + M_i(a) \, \lambda.
$$
The parameter $\lambda$ then corresponds to shifting some of the causal effect of $m$ into an association between $M_i(a)$ and $Y_i$, suggesting the existence of an unobserved confounder which, if unaccounted for, biases $\beta^Y_M$ by $\lambda$. In the absence of subject-matter expertise about the likely values of $\lambda$, one can use weak prior knowledge about the magnitude of the mediation effect to narrow down the range of plausible values. In the JOBS II study, we do this by constructing a pilot estimate of $r_y(m,a,x)$ of the form $\logit \widehat r_y(m,a,x) = \widehat \beta^Y_0 + x^\top \widehat \beta^Y_X + a\, \widehat \beta^Y_A + m \, \widehat \beta^Y_M$. In most cases, we feel that it is reasonable to assume that the effect of unmeasured confounding will not dominate the causal effect associated to $m$, so that $\lambda \in [-\widehat\beta^Y_M, \widehat\beta^Y_M]$ is a conservative collection of plausible values of $\lambda$.

The parameter $\rho$ measures dependence in the mediator process $M_i(\cdot)$.
In our experience, $\rho$ exerts less of an influence on the results than $\lambda$; the only impact of $\rho$ is due to the non-linearity of the link function, and we show in Section~\ref{sec:sensitivity-on-the-linear-scale} that the term $\rho$ plays no role when we use the linear link $\E_\theta\{Y_i(a,m) \mid M_i(a), M_i(a'), X_i\} = r_y(m,a,x) + \lambda\{M_i(a) - m\}$ instead of the logit link.

\begin{remark}
  The use of copulas in SI2A bears a passing resemblance to our use of comonotone random variables to reduce Monte Carlo error in Section~\ref{sec:posterior-computation-and-inference}. In this case, however, the choice of $\rho$ \emph{does} impact the model, and so we can no longer reduce the Monte Carlo error by making $M_{ik}^\star(a)$ and $M_{ik}^\star(a')$ comonotone.
\end{remark}


\subsection{Sensitivity on the Linear Scale}
\label{sec:sensitivity-on-the-linear-scale}

Part of the motivation for introducing a sensitivity parameter through the logit link in Section~\ref{sec:sensitivity-on-the-logit-scale} is that it is range preserving in the sense that it ensures that $0 < \E_\theta\{Y_i(a,m)\} < 1$. An alternative approach is to allow for possible violations of the range-preserving property and instead introduce $\lambda$ through a linear link. We consider the following assumption to replace SI2:

\begin{itemize}
    \item[SI2C] Conditional on $X_i$, $M_i(0)$, and $M_i(1)$, the mean of $Y_i(a,m)$ is given by
    \begin{align*}
        \E_\theta\{Y_i(a,m) \mid M_i(a), M_i(a'), X_i\}
        =
        r_y(m,a,x) + \lambda \{M_i(a) - m\}
    \end{align*}
    where $r_y(m,a,x) = \E_\theta(Y_i \mid M_i = m, A_i = a, X_i = x)$.
\end{itemize}

The following result establishes that SI2C identifies the average causal mediation effects. It is proved in Appendix~\ref{sec:proofs}.

\begin{proposition}
  \label{prop:si2c}
  Suppose that SI1, SI2C, and SI3 hold. Then we have
  \begin{align*}
    \E_\theta[Y_i\{a, M_i(a')\}]
    &= 
    \iint r_y(m,a,x) \, f_\theta(M_i = m \mid A_i = a', X_i = x) \ dm \ F_X(dx)
    \\&\qquad+
    \lambda [\E_\theta\{M_i(a)\} - \E_\theta\{M_i(a')\}]
    .
  \end{align*}
\end{proposition}

SI2C has several advantages over SI2A --- SI2B. First, we feel that shifts directly on the scale of the mean are more easily interpreted than shifts on the logit scale.
Like the shift on the logit scale developed in Section~\ref{sec:sensitivity-on-the-logit-scale}, $\lambda$ can be interpreted as shifting the causal effect of $m$ into an association with $M_i(a)$. For example, if we had used the linear model $r_y(m,a,x) = \beta_0^Y + x^\top \beta_X^Y + a \, \beta_A^Y + m \, \beta^Y_M$ then we could rewrite $\E_\theta\{Y_i(a, m) \mid M_i(a), M_i(a'), X_i\}$ as $\beta_0^Y + x^\top \beta^Y_X + a\, \beta_A^Y + m \, (\beta_M^Y - \lambda) + M_i(a) \, \lambda$.  To a elicit a default range of $\lambda$'s we can use the approach outlined in Section~\ref{sec:sensitivity-on-the-logit-scale} by fitting a linear regression of $Y_i$ on $(M_i, A_i, X_i)$.

The second benefit of SI2C is that the linearity removes the need to specify the correlation $\rho$ between $M_i(a)$ and $M_i(a')$ so that the sensitivity analysis only requires eliciting a range of plausible $\lambda$'s. This is very helpful, as $\rho$ is more difficult to interpret than $\lambda$.

Using Proposition~\ref{prop:si2c} we can again develop a Monte Carlo implementation of the $g$-formula using the approximation
\begin{math}
    \E_\theta[Y_{i}\{a,M_i(a')\}]
    \approx
    K^{-1} \sum_{i,k} \omega_i \, Y^\star_{ik}\{a, M_{ik}^\star(a')\}, 
\end{math}]
where $Y^\star_{ik}\{a,M_{ik}^\star(a')\} = r_y\{M_{ik}^\star(a'), a, X_i\} + \lambda \{M_{ik}^\star(a) - M_{ik}^\star(a')\}$ and $M_{ik}^\star(a)$ is sampled from $f_\theta(M_i \mid A_i = a, X_i)$. A possible algorithm is given in Algorithm~\ref{alg:linear} of the Supplementary Material.

\section{Illustrations}
\label{sec:illustrations}

\subsection{Application to JOBS II}
\label{sec:application}

We apply our methodology to the JOBS II dataset using SI as a benchmark. As potential confounders we include numeric variables measuring economic hardship (\texttt{econ\_hard}), baseline depression (\texttt{depress1}), and age (\texttt{age}), as well as the categorical variables measuring (\texttt{sex}), race (\texttt{nonwhite}), income bracket (\texttt{income}), occupation (\texttt{occp}), martial status (\texttt{martial}), and education level (\texttt{educ}). 

We posit generalized linear models for each component of the ZOIB models with a homogeneous effect for the treatment and mediator. That is, we use linear predictors of the form $X_i^\top\beta^M + \theta^M \, A_i$ for the mediator and linear predictors of the form $X_i^\top\beta^Y + \theta^Y \, A_i + \eta \, M_i$ for the outcome (with separate coefficients for each model component). Note that the homogeneity assumption implies neither that $\delta(0) = \delta(1)$ nor that $\zeta(0) = \zeta(1)$ due to the nonlinearity of the link functions in \eqref{eq:glms}.

\begin{figure}
    \centering
    \includegraphics[width=1\textwidth]{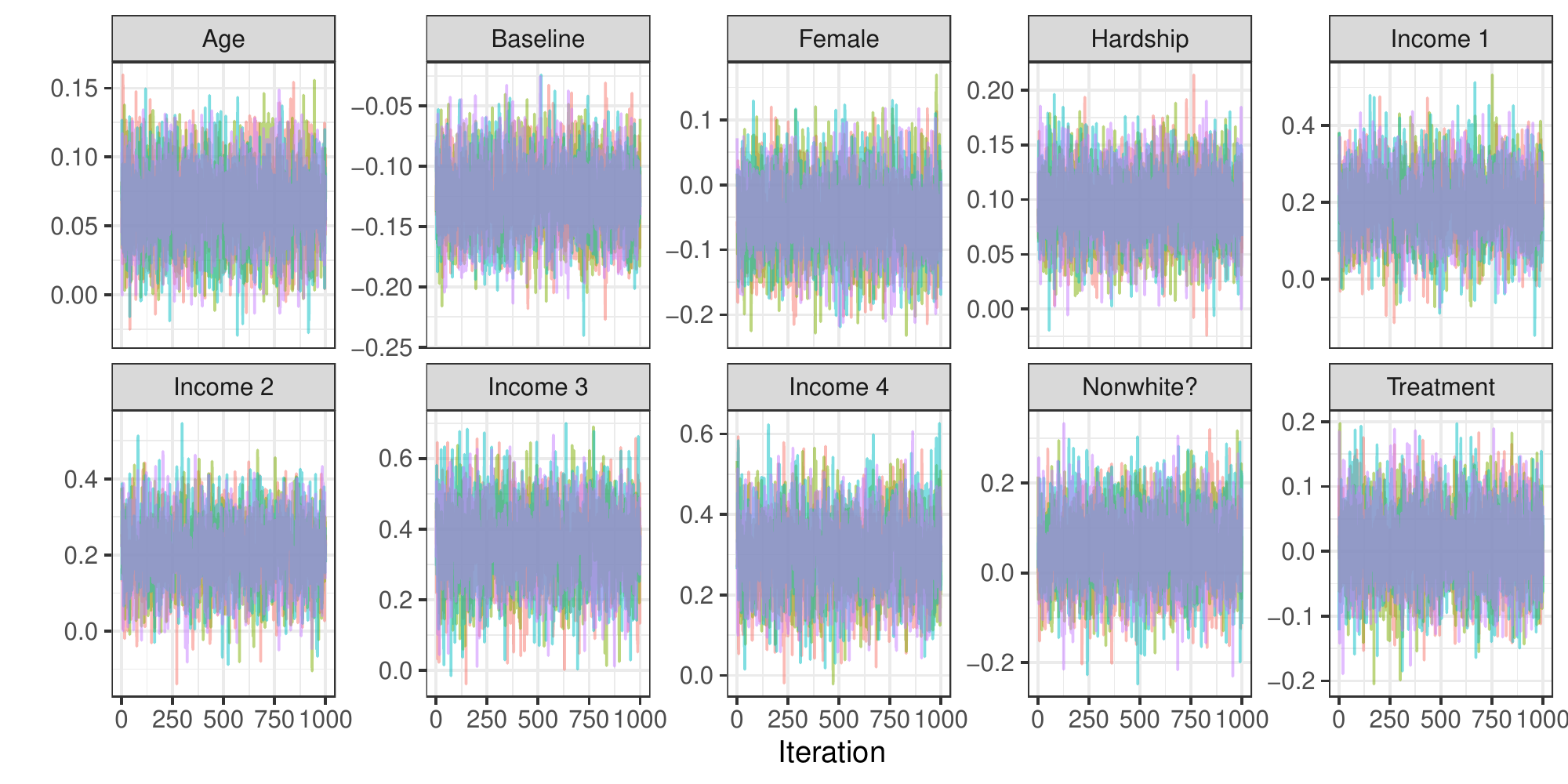}
    \caption{
      Mixing of our four Markov chains for the regression coefficients $\beta_\mu^M$ after discarding the first 1000 iterations to burn-in. Results are similar for the other sets of regression coefficients. Baseline denotes the baseline depression level, Hardship denotes the numeric measure of economic hardship, and Income 1 --- Income 4 denote indicator variables for different income levels.
    }
    \label{fig:Mixing}
\end{figure}

\begin{figure}
    \centering
    \includegraphics[width=1\textwidth]{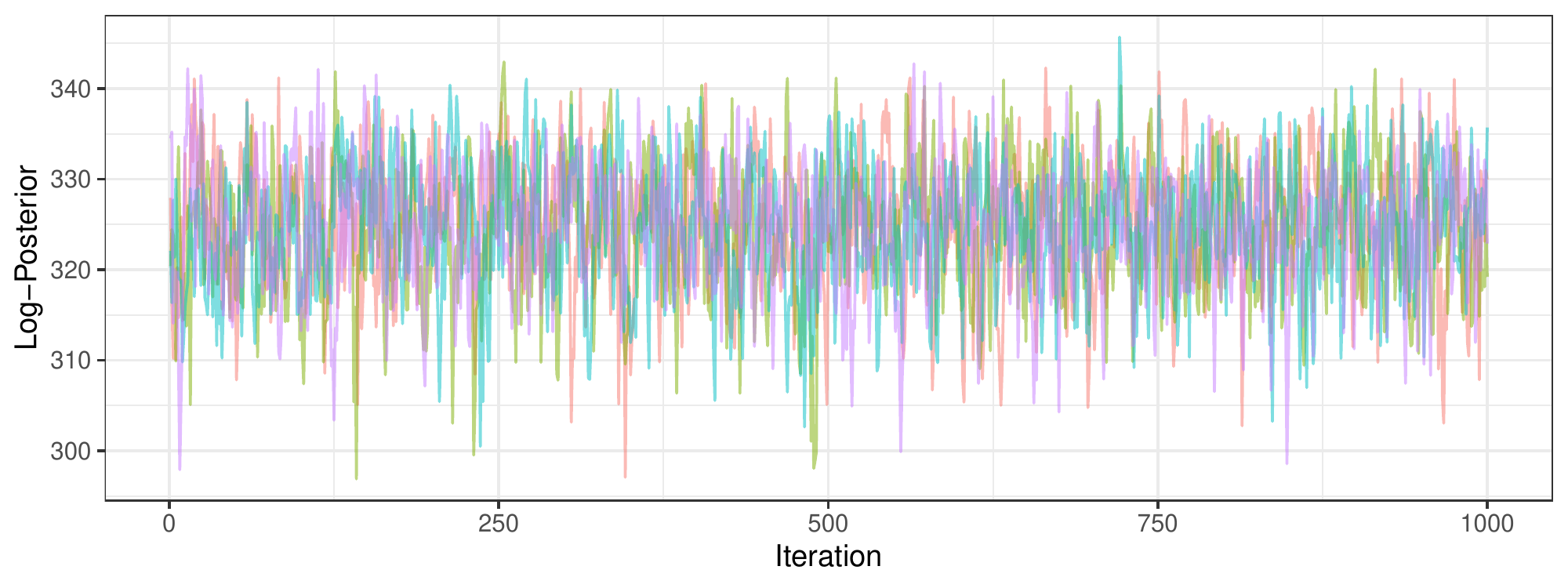}
    \caption{Mixing of the log-posterior density across the four chains.}
    \label{fig:Log-Posterior}
\end{figure}

The observed data models for $M_i$ and $Y_i$ are fit using Markov chain Monte Carlo (MCMC) in \texttt{Stan}, with a total of $8000$ samples collected over four parallel chains and $4000$ samples discarded to burn-in. There is no evidence of failure of the chains to mix: all traceplots indicate rapid mixing (see Figure~\ref{fig:Mixing} and Figure~\ref{fig:Log-Posterior}), all values of the Gelman-Rubin diagnostic \citep{gelman1992inference} $\widehat R$ are effectively $1$ (minimum of 0.9991, maximum of 1.003), and the minimal effective sample size across all of the monitored parameters is 1807.

\begin{figure}
    \centering
    \includegraphics[height=0.4\textheight]{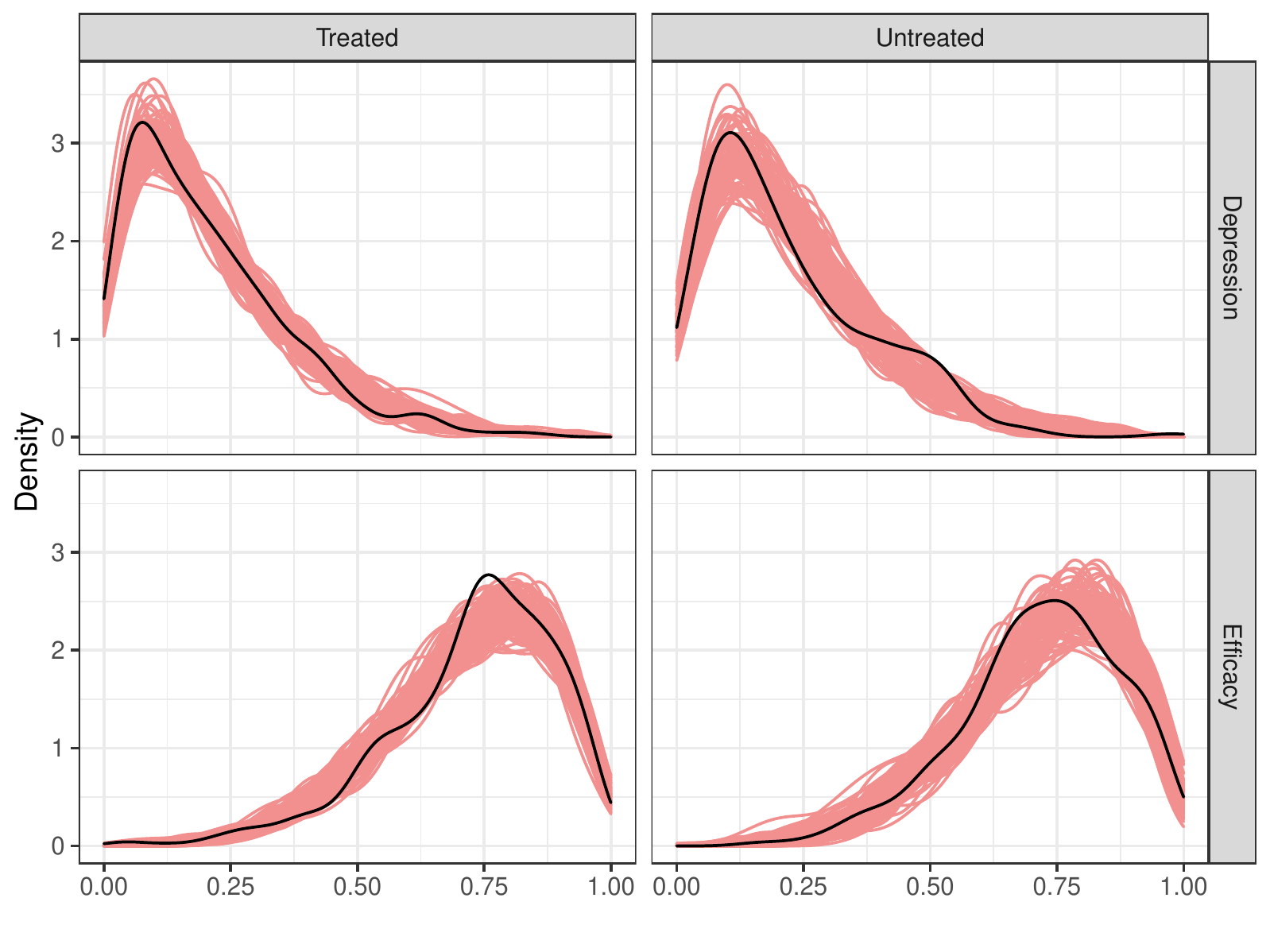}
    \caption{Kernel density estimates of the non-boundary proportion of the original data (black) and 100 replicated datasets (red).}
    \label{fig:Lexi-Figure-05}
\end{figure}

We now perform posterior predictive checks to compare the observed data $Y_i$ and $M_i$ to replicate datasets simulated from the fitted model. The goal of these checks is to assess how well the fitted model aligns with the observed data. In Figure~\ref{fig:Lexi-Figure-05} we check the fit of the beta distribution to the continuous part of the observed mediator/outcome distributions by comparing a kernel density estimate of the observed-data distribution of $Y_i$ and $M_i$ to 100 replicated datasets sampled from the posterior predictive distribution; the top row shows density estimates for the depression level under each treatment level for the continuous part of the data, while the bottom row shows the same for job-search self-efficacy. The posterior predictive distribution produces datasets that closely match the observed data, suggesting that the beta model for the continuous part of the data is adequate.

\begin{figure}
    \centering
    \includegraphics[height=0.4\textheight]{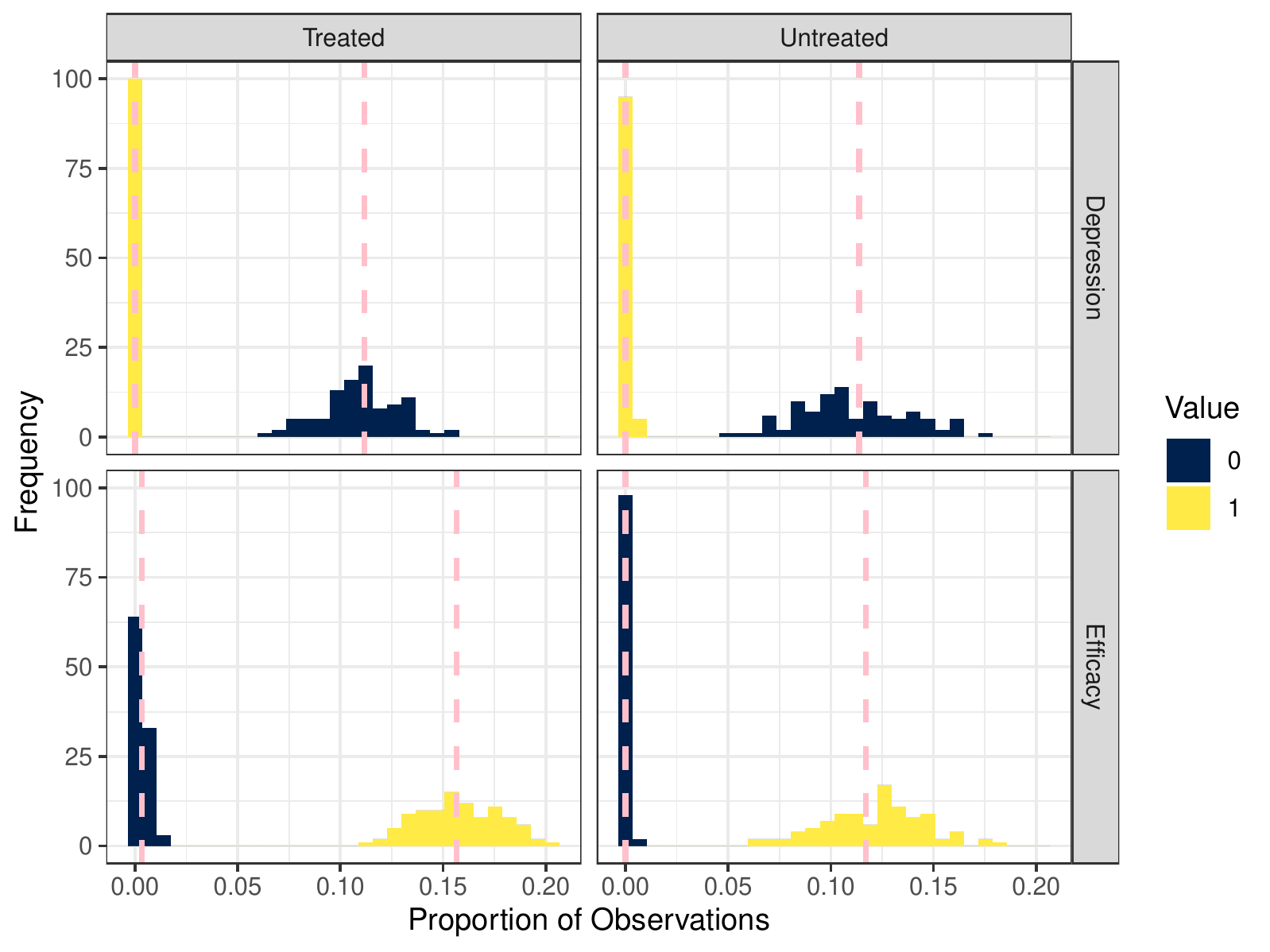}
    \caption{Histograms of the proportions of individuals taking the boundary values $0$ (in dark blue) and $1$ (in light yellow) across 100 replicated datasets for \comment{each of }the outcome (depression, first row) and mediator (efficacy, second row), separately for the treated and untreated groups. The proportions of the observed data taking boundary values are given by the pink vertical dashed lines.}
    \label{fig:Lexi-Figure-06}
\end{figure}

In Figure~\ref{fig:Lexi-Figure-06} we check the fit of the logistic regression models to the boundary points $0$ and $1$ by comparing the observed proportion of $0$s and $1$s for the outcome and mediator to what is observed in replicated datasets. Again there is close agreement between the model and simulated data.

Results for the average causal mediation effects are given in Table~\ref{tab:effects}. Summarizing, there is little evidence for either a direct or indirect effect of the treatment on the outcome; in particular, the signs of the mediation effects are uncertain. The story is similar for the quantile mediation effects.


\begin{table}
\centering
\begin{tabular}{lrrrrrr}
  \toprule
  Effect & Est. & SD & Lower & Upper & $Z$-Score & $P$-value \\ 
  \midrule
  $\delta(0)$ & -0.0110 & 0.0108 & -0.0330 & 0.0098 & -1.0131 & 0.3110 \\ 
  $\delta(1)$ & -0.0102 & 0.0101 & -0.0308 & 0.0089 & -1.0144 & 0.3104 \\ 
  $\zeta(0)$ & -0.0282 & 0.0403 & -0.1065 & 0.0491 & -0.7000 & 0.4839 \\ 
  $\zeta(1)$ & -0.0275 & 0.0400 & -0.1058 & 0.0490 & -0.6880 & 0.4915 \\ 
  $\tau$ & -0.0385 & 0.0416 & -0.1202 & 0.0415 & -0.9244 & 0.3553 \\ 
  \bottomrule
\end{tabular}
\caption{
  Effect estimates for the JOBS II data using the ZOIB formulation for the outcome (depression) and mediator (efficacy) assuming sequential ignorability. \label{tab:effects}
}
\end{table}

\subsection{Sensitivity Analysis}

While there is no evidence for either a direct or indirect treatment effect under SI, we may be concerned that the effects are being masked by unobserved confounding. We now apply the sensitivity analysis techniques introduced in Section~\ref{sec:sensitivity-analysis} to assess the impact of unmeasured confounding. 

\begin{figure}[t]
    \centering
    \includegraphics[height=0.43\textheight]{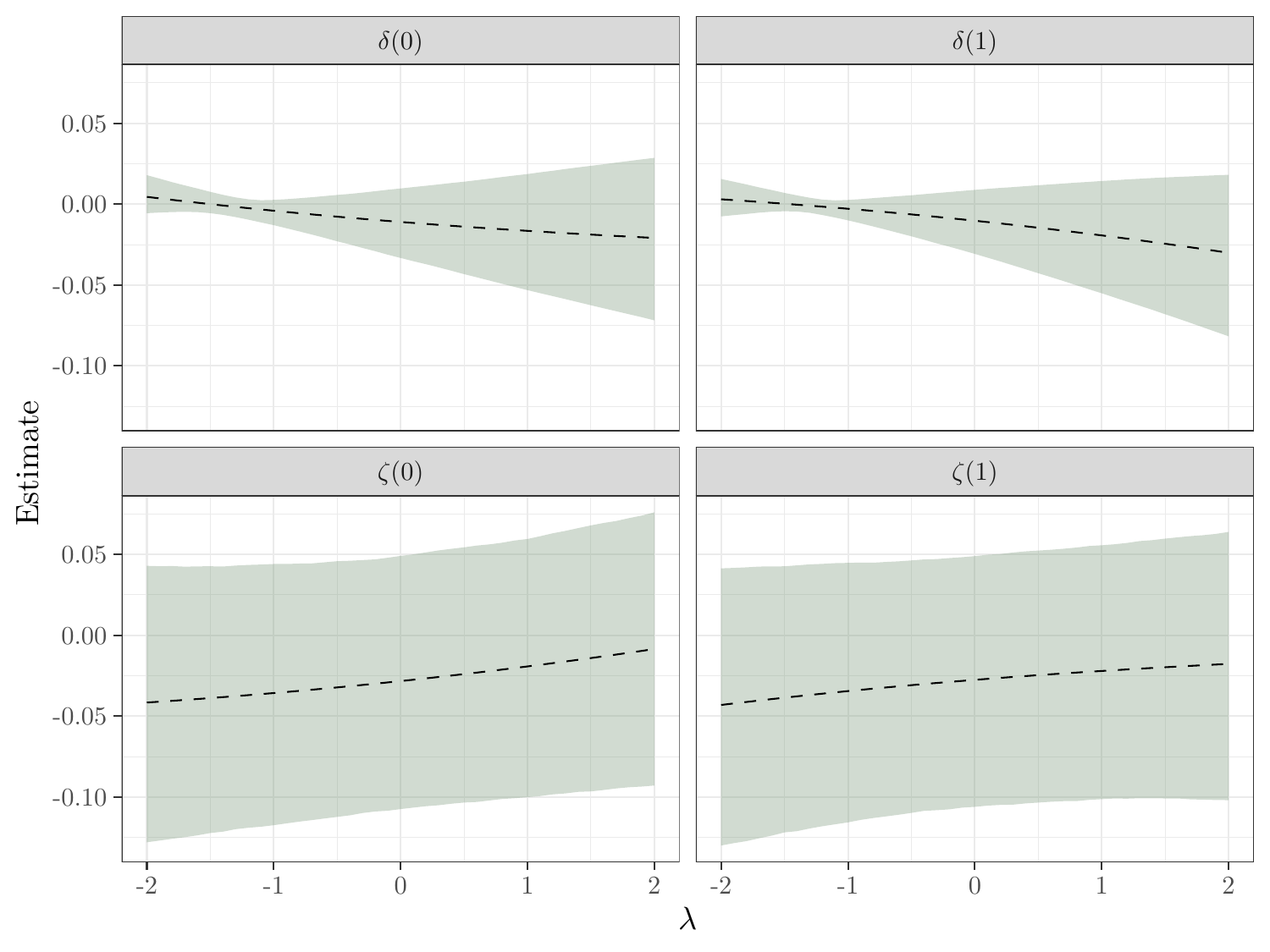}
    \caption{Sensitivity of inferences about $\delta(a)$ and $\zeta(a)$ to changes in the sensitivity parameter $\lambda$ under assumptions SI2A and SI2B. The dashed line is the posterior mean, and the bands delimit a pointwise 95\% credible interval.}
    \label{fig:SensitivityPlot}
\end{figure}

We first consider the assumptions SI2A and SI2B. Using the fixed value and large $\rho = 0.95$, Figure~\ref{fig:SensitivityPlot} shows how inferences about the mediation effects change as $\lambda$ is varied across a range of plausible values. To calibrate $\lambda$, we fit a linear model to the conditional mean $\logit \E(Y_i \mid M_i, A_i, X_i)$ using quasi-likelihood and then considered values of $\lambda$ no more than twice as large in magnitude than the estimated effect of $M_i$ in this model; this corresponds to the belief that most of the association of $Y_i$ with $M_i$ should be attributable to a causal effect of the mediator rather than confounding between the outcome and mediator processes.

From Figure~\ref{fig:SensitivityPlot}, inferences for the direct effects $\zeta(a)$ are robust to unmeasured confounding between $Y_i(a,m)$ and $M_i(a)$, while inferences for the indirect effects $\delta(a)$ are less robust. When $\lambda$ is negative there is less uncertainty in $\delta(a)$, although the estimates are also pulled towards zero. For larger values of $\lambda$, the estimates of $\delta(a)$ are larger (in magnitude), but also more uncertain. The substantive conclusion remains the same: there is little evidence regarding the sign of either the direct or indirect effects.

\begin{figure}[t]
    \centering
    \includegraphics[height=0.43\textheight]{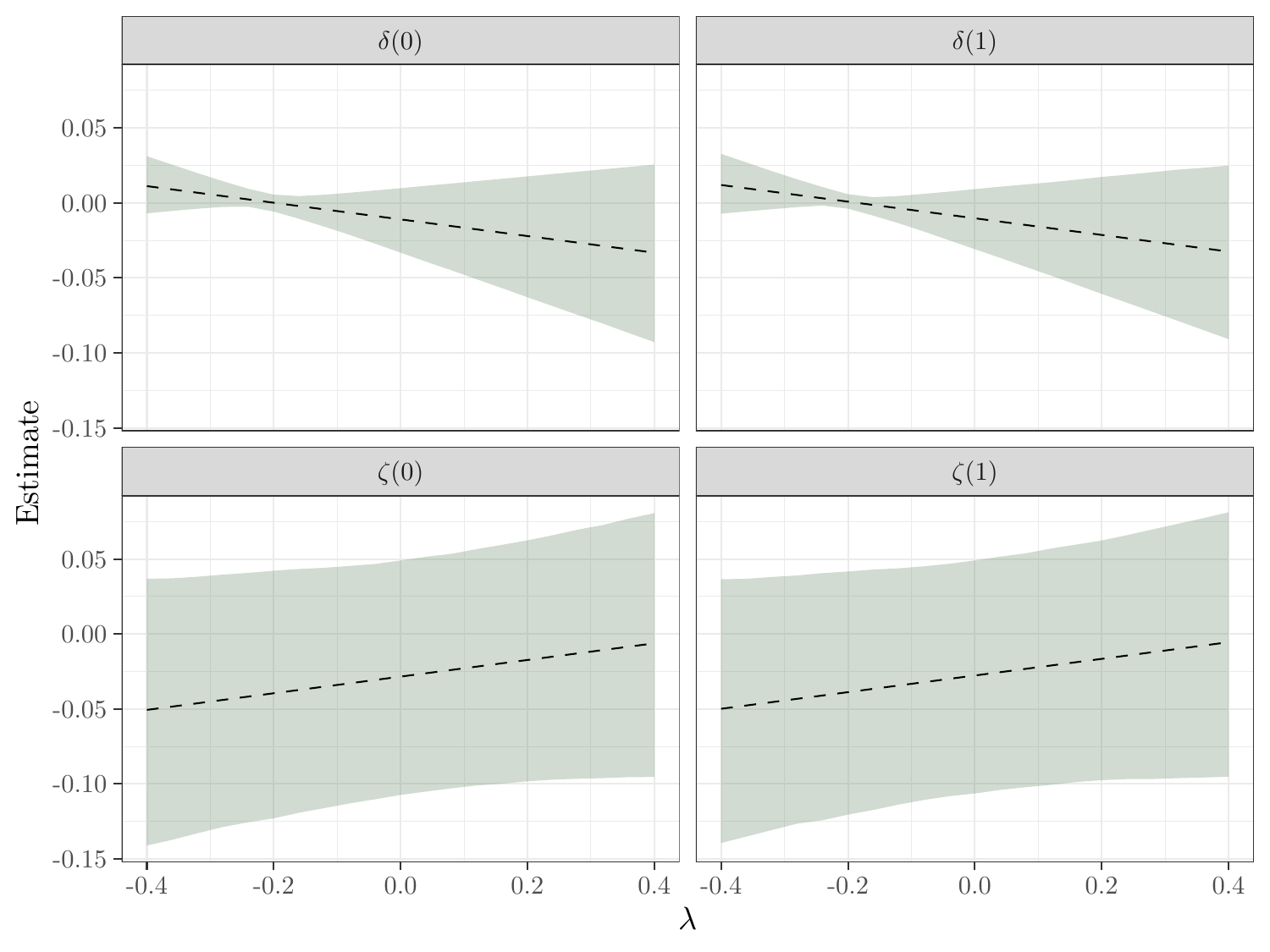}
    \caption{Sensitivity of inferences about $\delta(a)$ and $\zeta(a)$ to changes in the sensitivity parameter $\lambda$ under assumption SI2C. The dashed line is the posterior mean, and the bands delimit a pointwise 95\% credible interval.\label{fig:SensitivityPlotLinear}}
\end{figure}

Figure~\ref{fig:SensitivityPlotLinear} gives a similar sensitivity analysis under SI2C, with a reasonable range for $\lambda$ now obtained from a linear regression of $Y_i$ on $(A_i, M_i, X_i)$. The results are substantively in agreement with the logit-scaled sensitivty analysis: there are no values of $\lambda$ that lead to evidence of either direct or indirect effects of the treatment. 

For the linear model, it is easy to see why $\lambda$ does not greatly influence our conclusions: as shown in Proposition~\ref{prop:si2c}, the influence of $\lambda$ is through the average treatment effect on the mediator $\varpi = \E_\theta(M_i \mid A_i = 1) - \E_\theta(M_i \mid A_i = 0)$. Because the sign of $\varpi$ is uncertain, the direction by which $\lambda$ shifts $\E_\theta[Y_i\{a, M_i(a')\}]$ is itself uncertain, so that increasing $|\lambda|$ has more effect on the uncertainty of the mediation effects than it does on the point estimates.

\subsection{Simulation Example}
We evaluate Algorithm~\ref{alg:zoib} under a variety of different data generating mechanisms based on the JOBS II data. To devise relevant simulation settings, we first fit our model to the JOBS II data and then modified the estimated coefficients of the fitted beta-regression model. We consider homogeneous effects for both the mediator and outcome, and write $\xi^M$ and $\xi^Y$ for the estimated coefficient for the effect of $A_i$ on $M_i$ and $Y_i$ respectively. The following features of the data generating mechanism were varied.

\begin{description}[leftmargin=0em, style = unboxed]
  \item[Sample Size] We consider $N \in \{899, 1798\}$, which is equal to the size of the JOBS II dataset and twice the size of the JOBS II dataset.
  \item[Treatment Effect on Mediator] We consider $\xi^M \in \{0,\widehat \xi^M,10\widehat \xi^M\}$; the first setting corresponds to no indirect effect of treatment, the second to a realistic indirect effect of the treatment, and the last to a very large indirect effect on the treatment.
  \item[Direct Treatment Effect on the Outcome] We consider $\xi^Y \in \{0,\widehat\xi^Y,10\widehat\xi^Y\}$, the choices of which are analogous to the ones for the treatment effect on the mediator.
\end{description}

We consider five data generating mechanisms: (1) no mediation, where $(\xi^Y, \xi^M) = (\widehat\xi^Y, 0)$; (2) complete mediation, where where $(\xi^Y, \xi^M) = (0, \widehat \xi^M)$; (3) strong no mediation, where $(\xi^Y, \xi^M) = (10 \widehat \xi^Y, 0)$; (4) strong complete mediation, where $(\xi^Y, \xi^M) = (0, 10\widehat \xi^M)$; and (5) no modifications, where $(\xi^Y, \xi^M) = (\widehat \xi^Y, \widehat \xi^M)$. For each scenario we used simulated 200 datasets to compute the bias of the effect estimates, the root mean-squared error (RMSE), coverage of nominal 95\% credible intervals, and the average length of a nominal 95\% interval. 



Table~\ref{tab:simulations} summarizes the results for each scenario. For readability, all entries of the table are multiplied by $100$.     Prior to the simulation experiment, we computed the true direct, indirect, and total effects using Monte Carlo integration with $90,799$ samples (101 times the size of the original data). For each simulated dataset we collected a total of 2000 samples across eight parallel chains, with 250 burn-in samples per chain. Our method performs well in terms of bias for both sample sizes and, as expected, we observe lower RMSEs for the larger sample size. The 95\% credible intervals are slightly conservative, particularly for the indirect effect; across all scenarios and effects, the smallest coverage probability was 94\%. Ultimately, the results show that our approach to computing the mediation effects, while tending to be conservative, appears to work well.


\begin{table}
\centering
\hspace*{-2.3em}
\begin{tabular}{ccrrrrrrrrr}
  \toprule
  & & & \multicolumn{4}{c}{$n = 899$} & \multicolumn{4}{c}{$n = 1798$}\\ \cmidrule(lr){4-7} \cmidrule(lr){8-11}
Scenario & Effect & Truth & Bias & RMSE & Coverage  & Length  & Bias & RMSE & Coverage  & Length  \\ 
  \midrule
    & $\delta(0)$ & -0.26 & 0.01 & 0.31 & 99.00 & 1.53 & 0.03 & 0.22 & 99.00 & 1.07 \\ 
    & $\delta(1)$ & -0.26 & 0.02 & 0.29 & 100.00 & 1.45 & 0.03 & 0.21 & 99.50 & 1.02 \\ 
  1 & $\zeta(0)$ & -0.98 & 0.00 & 1.06 & 97.50 & 4.26 & 0.06 & 0.72 & 95.50 & 3.02 \\ 
    & $\zeta(1)$ & -0.97 & 0.01 & 1.05 & 97.00 & 4.23 & 0.07 & 0.72 & 95.50 & 3.01 \\ 
    & $\tau$ & -1.23 & 0.02 & 1.11 & 96.50 & 4.41 & 0.09 & 0.74 & 95.50 & 3.13 \\ 
    \hline
    & $\delta(0)$ & -0.41 & 0.10 & 0.33 & 97.50 & 1.55 & 0.02 & 0.22 & 98.50 & 1.08 \\ 
    & $\delta(1)$ & -0.40 & 0.09 & 0.32 & 98.00 & 1.53 & 0.01 & 0.22 & 98.50 & 1.06 \\ 
  2 & $\zeta(0)$ & 0.22 & 0.01 & 1.11 & 95.00 & 4.33 & 0.01 & 0.73 & 95.50 & 3.06 \\ 
    & $\zeta(1)$ & 0.24 & 0.00 & 1.10 & 95.00 & 4.30 & 0.00 & 0.73 & 95.00 & 3.03 \\ 
    & $\tau$ & -0.17 & 0.10 & 1.18 & 94.50 & 4.49 & 0.02 & 0.80 & 95.00 & 3.17 \\ 
       \hline
    & $\delta(0)$ & -0.11 & -0.11 & 0.30 & 98.50 & 1.49 & -0.13 & 0.26 & 96.50 & 1.05 \\ 
    & $\delta(1)$ & -0.10 & -0.02 & 0.16 & 99.50 & 0.93 & -0.04 & 0.12 & 99.00 & 0.65 \\ 
  3 & $\zeta(0)$ & -9.75 & 0.06 & 1.03 & 95.50 & 3.98 & -0.12 & 0.74 & 94.50 & 2.81 \\ 
    & $\zeta(1)$ & -9.73 & 0.14 & 1.03 & 94.00 & 3.96 & -0.02 & 0.73 & 95.50 & 2.79 \\ 
    & $\tau$ & -9.84 & 0.03 & 1.04 & 94.50 & 4.08 & -0.15 & 0.75 & 94.00 & 2.88 \\ 
      \hline
    & $\delta(0)$ & -1.34 & 0.05 & 0.38 & 96.00 & 1.64 & 0.06 & 0.28 & 97.50 & 1.14 \\ 
    & $\delta(1)$ & -1.37 & 0.09 & 0.38 & 94.00 & 1.64 & 0.11 & 0.29 & 97.00 & 1.14 \\ 
  4 & $\zeta(0)$ & 0.29 & 0.02 & 1.11 & 95.00 & 4.37 & -0.07 & 0.72 & 97.00 & 3.09 \\ 
    & $\zeta(1)$ & 0.26 & 0.06 & 1.08 & 94.50 & 4.25 & -0.02 & 0.70 & 97.50 & 3.01 \\ 
    & $\tau$ & -1.08 & 0.11 & 1.13 & 94.50 & 4.43 & 0.04 & 0.74 & 96.50 & 3.14 \\ 
      \hline
    & $\delta(0)$ & -0.33 & 0.02 & 0.33 & 98.50 & 1.53 & 0.02 & 0.22 & 99.50 & 1.08 \\ 
    & $\delta(1)$ & -0.34 & 0.06 & 0.31 & 98.50 & 1.44 & 0.05 & 0.21 & 99.00 & 1.01 \\ 
  5 & $\zeta(0)$ & -0.99 & -0.07 & 0.95 & 96.50 & 4.28 & -0.07 & 0.74 & 95.00 & 3.03 \\ 
    & $\zeta(1)$ & -1.00 & -0.04 & 0.94 & 96.50 & 4.24 & -0.03 & 0.73 & 95.00 & 3.01 \\ 
    & $\tau$ & -1.33 & -0.01 & 1.05 & 96.50 & 4.42 & -0.02 & 0.78 & 96.50 & 3.13 \\
   \bottomrule
\end{tabular}
\caption{
  JOBS II simulation results for average mediation effects using Algorithm~\ref{alg:zoib}. Each value in the table is multiplied by 100. The columns correspond to bias, the true values of the effect, the root-mean-squared error (RMSE), the coverage of nominal 95\% credible intervals, and the average length of a nominal 95\% credible interval. Scenarios 1-5 correspond to the following: (1) $\xi^M = 0$ and $\xi^Y = \widehat \xi^Y$; (2) complete mediation, where $(\xi^Y, \xi^M) = (0, \widehat \xi^M)$; (3) $\xi^M = 0$ and $\xi^Y = 10$; (4) complete mediation, where $(\xi^Y, \xi^M) = (0, 10\widehat \xi^M)$; (5) no modifications, i.e., $(\xi^Y, \xi^M) = (\widehat \xi^Y, \widehat \xi^M)$, where $\xi^M$ and $\xi^Y$ are the estimated coefficients for the effect of $A_i$ on $M_i$ and $Y_i$, respectively. \label{tab:simulations}
}
\end{table}

\section{Discussion}
\label{sec:discussion}

In many practical situations, either the mediator or outcome (or both) will have a mixed-scale distribution, necessitating the use of models beyond the usual linear and generalized linear models. We proposed a zero-one inflated beta distribution after scaling the data to lie in $[0,1]$. This family of distributions includes many distributional shapes.

The framework proposed here is flexible enough for users to adapt Algorithm~\ref{alg:gc} and Algorithm~\ref{alg:gc-quantile} to fit essentially arbitrary zero-one inflated models; for example, it is straight-forward to adapt this approach to handle Tobit regression models \citep{mcdonald1980uses}. It is also straight-forward to extend our approach to a Bayesian nonparametric setting using Dirichlet process mixture models \citep{kim2016framework} and nonparametric Bayesian additive regression tree models \citep{li2020adaptive}.

Perhaps just as important as the model we have proposed is our description of how to perform a sensitivity analysis with this type of data. The framework for sensitivity analysis we presented is simple, interpretable, and also extends easily to other model specifications such as the Tobit model.

While we have taken a Bayesian approach in this paper, there is nothing in principle which stops us from applying our approach in the Frequentist framework. As shown by \citet{linero2021simulation}, the $g$-computation framework we used here can be used with the nonparametric bootstrap as well; hence, we could have performed maximum likelihood estimation and used the bootstrap to perform uncertainty quantification rather than using Bayesian inference.

Code and a package which replicates our analysis is available online at \url{www.github.com/theodds/ZOIBMediation}.

\appendix

\section{Proofs}
\label{sec:proofs}

\subsection{Proof of Proposition~\ref{prop:logit}}

By iterated expectation and SI2B we have
\begin{align*}
  \E_\theta[Y_i\{a, M_i(a')\}]
  &=
  \E_\theta \E_\theta[Y_i\{a, M_i(a')\} \mid M_i(a), M_i(a'), X_i]
  \\&=
  \E_\theta \expit[\widetilde r_y\{M_i(a'), a, X_i\} + \lambda\{M_i(a) - M_i(a')\}].
\end{align*}
By SI2A, we have that $M_i(a) = F_M^-\{\Phi(Z_{ia}) \mid A_i = a, X_i = x\}$ and $M_i(a') = F_M^-\{\Phi(Z_{ia'}) \mid A_i = a', X_i = x\}$ where $(Z_{i0}, Z_{i1})$ is jointly standard normal with correlation matrix $\Sigma$. Therefore
\begin{align*}
    &\E_\theta \expit[\widetilde r_y\{M_i(a'), a, X_i\} + \lambda\{M_i(a) - M_i(a')\}]
    \\&\quad
    = 
    \iiint
    \expit\{\widetilde r_y(m',a,x) + \lambda(m - m')\}
    \, 
    \Normal\{(z_0, z_1)^\top \mid (0,0)^\top, \Sigma\}
    \ dz_0 \ dz_1 \ F_X(dx)
\end{align*}
where $m \equiv F_M^-(z_a \mid A_i = a, X_i = x)$ and $m' = F_M^-(z_{a'} \mid A_i = a', X_i = x)$, completing the proof.

\subsection{Proof of Proposition~\ref{prop:si2c}}

As in the proof of Proposition~\ref{prop:logit}, iterated expectation gives
\begin{align}
    \label{eq:starting-point}
    \E_\theta[Y_i\{a, M_i(a')\}]
    =
    \E_\theta[r_y\{M_i(a'), a, X_i\} + \lambda\{M_i(a) - M_i(a')\}].
\end{align}
so that
\begin{math}
    \E_\theta[Y_i\{a, M_i(a')\}]
    =
    \E_\theta[r_y\{M_i(a'), a, X_i\}] + 
    \lambda[\E_\theta\{M_i(a)\} - \E_\theta\{M_i(a')\}].
\end{math}
An application of SI1 gives
\begin{align*}
    \E_\theta[r_y\{M_i(a'), a, X_i\}]
    &=
    \iint r_y(m,a,x) \, f_\theta\{M_i(a') = m \mid X_i = x\} \ dm \ F_X(dx) 
    \\&= \int r_y(m,a,x) \, f_\theta\{M_i(a') = m \mid A_i = a', X_i = x\} \ dm \  F_X(dx) 
    \\&= \iint r_y(m,a,x) \, f_\theta(M_i = m \mid A_i = a', X_i = x) \ dm \  F_X(dx).
\end{align*}
Plugging this expressions for $\E_\theta[r_y\{M_i(a'), a, X_i\}]$ into \eqref{eq:starting-point} finishes the proof.

\bibliographystyle{apalike}
\bibliography{mybib}

\clearpage

\singlespacing
\makeatletter
\def\@thanks{}
\makeatother
\setlength{\headheight}{15pt}

\title{Supplementary Material to Causal Mediation and Sensitivity Analysis for Mixed-Scale Data}


\maketitle

\doublespacing
\renewcommand{\thesection}{S.\arabic{section}}
\renewcommand{\theequation}{S.\arabic{equation}}
\renewcommand{\thepage}{S.\arabic{page}}
\renewcommand{\thealgorithm}{S.\arabic{algorithm}}
\setcounter{page}{1}
\setcounter{section}{0}
\setcounter{algorithm}{0}
\thispagestyle{fancy}
\pagestyle{fancy}
\rhead{\textbf{Not for Publication Supplementary Material}}

\section{Algorithms}

In this section we provide algorithms for average mediation effect estimation using the sensitivity analysis techniques developed in Section~\ref{sec:sensitivity-analysis}. Algorithm \ref{alg:sensitivity} implements the logit-scale sensitivity analysis of Section~\ref{sec:sensitivity-on-the-logit-scale} while Algorithm~\ref{alg:linear} implements the linear-scale sensitivity analysis of Section~\ref{sec:sensitivity-on-the-linear-scale}.

\begin{algorithm}
  \caption{Monte Carlo $g$-formula With Sensitivity Parameters\label{alg:sensitivity}}
  \textbf{Input:} $\theta, K, \rho, \lambda, \{X_i\}_{i=1}^N$
    \begin{algorithmic}[1]
    \State{Sample $\omega \sim \Dirichlet(1,\ldots,1)$}
    \For{$k = 1,\ldots,K$}
      \For{$i = 1,\ldots,N$}
        \State{\parbox[t]{.9\linewidth}{Sample $(U^\star_0, U^\star_1) \sim \Normal(\mathbf 0, \Sigma)$ where $\Sigma$ is a correlation matrix with correlation $\rho$.}}
        \State Set $U_0 \gets \Phi(U^\star_0)$ and $U_1 \gets \Phi(U^\star_1)$
        \For{$a = 0,1$}
          \State{\parbox[t]{.8\linewidth}{Compute \newline
            $\logit \alpha^M \gets X_i^\top \beta^M_\alpha(a)$,
            $\logit \gamma^M \gets X_i^\top \beta^M_\gamma(a)$, \newline
            $\logit \mu^M \gets X_i^\top \beta^M_\mu(a)$,
            $\log \phi^M \gets X_i^\top \beta^M_\phi(a)$.}
          }
          \State{
            Set
            \begin{align*}
              M_{ik}^\star(a)
              \gets
              \begin{cases}
                0 \qquad & \text{if $U_a < \alpha^M$}, \\
                1 \qquad & \text{if $U_a > 1 - (1-\alpha^M)\gamma^M$}, \\
                F_{\Beta}^{-1}\{U_a' \mid \mu^M \phi^M, (1-\mu^M)\phi^M\} \qquad & \text{otherwise}
              \end{cases}
            \end{align*}
            \hspace*{4.8em}where $U_a' = (U_a - \alpha^M) / [(1 - \alpha^M)(1 - \gamma^M)]$.
          }
          \For{$a' = 0,1$}
            \State{\parbox[t]{0.83\linewidth}{Compute \newline
        $\logit \alpha^Y \gets (X_i, M_{ik}^\star(a))^\top \beta^Y_\alpha(a')$,
        $\logit \gamma^Y \gets (X_i, M_{ik}^\star(a))^\top \beta^Y_\gamma(a')$, \newline
        $\logit \mu^Y \gets (X_i, M_{ik}^\star(a))^\top \beta^Y_\mu(a')$,
        $\log \phi^Y \gets (X_i, M_{ik}^\star(a))^\top  \beta^Y_\phi(a')$.
            }}
            \State{Set $E \gets (1 - \alpha^Y) \gamma^Y + (1-\alpha^Y)(1-\gamma^Y) \mu^Y$.}
            \State{Set $Y^\star_{ik}\{a', M^\star_{ik}(a)\} \gets \expit\{\logit E + \lambda \{M^\star_{ik}(a') - M^\star_{ik}(a)\}$.}
          \EndFor
        \EndFor
      \EndFor
    \EndFor
    \State{Approximate $\delta(a),\zeta(a), \tau$ with
      \begin{align*}
        \delta(a) & \approx K^{-1}\sum_{i,k}\omega_i [Y_{ik}^{\star}\{a,M_{ik}^{\star}(1)\}-Y_{ik}^{\star}\{a,M_{ik}^{\star}(0)\}] \\
        \zeta(a) & \approx K^{-1}\sum_{i,k}\omega_i[Y_{ik}^{\star}\{1,M_{ik}^{\star}(a)\}-Y_{ik}^{\star}\{0,M_{ik}^{\star}(a)\}]  \\
        \tau & \approx K^{-1}\sum_{i,k}\omega_i[Y_{ik}^{\star}\{1,M_{ik}^{\star}(1)\}-Y_{ik}^{\star}\{0,M_{ik}^{\star}(0)\}]
      \end{align*}   
      \hspace*{0em}for $a = 0,1$
    }
    \State{\Return{$\{\delta(0), \delta(1), \zeta(0), \zeta(1)\}$}}
  \end{algorithmic}
\end{algorithm}

\begin{algorithm}[t]
  \caption{Monte Carlo $g$-formula With Linear Sensitivity Parameters\label{alg:linear}}
  \textbf{Input:} $\theta, K, \lambda, \{X_i\}_{i=1}^N$
    \begin{algorithmic}[1]
    \State{Sample $\omega \sim \Dirichlet(1,\ldots,1)$}
    \For{$k = 1,\ldots,K$}
      \For{$i = 1,\ldots,N$}
        \State{Sample $U \sim \Uniform(0,1)$}
        \For{$a = 0,1$}
          \State{\parbox[t]{.83\linewidth}{Compute \newline
            $\logit \alpha^M \gets X_i^\top \beta^M_\alpha(a)$,
            $\logit \gamma^M \gets X_i^\top \beta^M_\gamma(a)$, \newline
            $\logit \mu^M \gets X_i^\top \beta^M_\mu(a)$,
            $\log \phi^M \gets X_i^\top \beta^M_\phi(a)$.}
          }
          \State{
            Set
            \begin{align*}
              M_{ik}^\star(a)
              \gets
              \begin{cases}
                0 \qquad & \text{if $U < \alpha^M$}, \\
                1 \qquad & \text{if $U > 1 - (1-\alpha^M)\gamma^M$}, \\
                F_{\Beta}^{-1}\{U' \mid \mu^M \phi^M, (1-\mu^M)\phi^M\} \qquad & \text{otherwise}
              \end{cases}
            \end{align*}
            \hspace*{4.8em}where $U' = (U - \alpha^M) / [(1 - \alpha^M)(1 - \gamma^M)]$.
          }
          \For{$a' = 0,1$}
            \State{\parbox[t]{0.83\linewidth}{Compute \newline
        $\logit \alpha^Y \gets (X_i, M_{ik}^\star(a))^\top \beta^Y_\alpha(a')$,
        $\logit \gamma^Y \gets (X_i, M_{ik}^\star(a))^\top \beta^Y_\gamma(a')$, \newline
        $\logit \mu^Y \gets (X_i, M_{ik}^\star(a))^\top \beta^Y_\mu(a')$,
        $\log \phi^Y \gets (X_i, M_{ik}^\star(a))^\top  \beta^Y_\phi(a')$.
            }}
            \State{Set $Y^\star_{ik}\{a', M_i(a)\} \gets (1 - \alpha^Y) \gamma^Y + (1 - \alpha^Y)(1-\gamma^Y) \mu^Y + \lambda \{M^\star_{ik}(a') - M^\star_{ik}(a)\}$}.
          \EndFor
        \EndFor
      \EndFor
    \EndFor
    \State{Approximate $\delta(a),\zeta(a), \tau$ with
      \begin{align*}
        \delta(a) & \approx K^{-1}\sum_{i,k}\omega_i [Y_{ik}^{\star}\{a,M_{ik}^{\star}(1)\}-Y_{ik}^{\star}\{a,M_{ik}^{\star}(0)\}] \\
        \zeta(a) & \approx K^{-1}\sum_{i,k}\omega_i[Y_{ik}^{\star}\{1,M_{ik}^{\star}(a)\}-Y_{ik}^{\star}\{0,M_{ik}^{\star}(a)\}]  \\
        \tau & \approx K^{-1}\sum_{i,k}\omega_i[Y_{ik}^{\star}\{1,M_{ik}^{\star}(1)\}-Y_{ik}^{\star}\{0,M_{ik}^{\star}(0)\}]
      \end{align*}   
      \hspace*{0em}for $a = 0,1$
    }
    \State{\Return{$\{\delta(0), \delta(1), \zeta(0), \zeta(1)\}$}}
  \end{algorithmic}
\end{algorithm}

\end{document}